\documentclass[
reprint,
superscriptaddress,
amsmath,amssymb,
aps, 
prb,
twocolumn,
]{revtex4-2}

\usepackage{graphicx}% Include figure files
\usepackage{dcolumn}% Align table columns on decimal point
\usepackage{bm}% bold math
\usepackage{xcolor} % for \textcolor
\usepackage{subcaption}
\usepackage[percent]{overpic} % add this in your preamble
\usepackage[font=small,labelfont=bf,justification=raggedright,singlelinecheck=false]{caption}
\usepackage[normalem]{ulem}
\usepackage{amsmath}
\usepackage{amssymb}

\def\opone{\leavevmode\hbox{\small1\kern-3.8pt\normalsize1}}

\begin{document}

\preprint{APS/123-QED}

\title{\textbf{Spectroscopy and Coherence of an Excited-State Transition in Tm\textsuperscript{3+}\!:\,YAlO\textsubscript{3}} at Telecommunication Wavelength}% 

\author{Luozhen Li}
 \altaffiliation{These authors contributed equally to this work.}%Lines break automatically or can be forced with \\
\affiliation{Department of Applied Physics, University of Geneva, 1211 Geneva, Switzerland}
\affiliation{Constructor University Bremen, 28759 Bremen, Germany} 

\author{Akshay Babu Karyath}
 \altaffiliation{These authors contributed equally to this work.}
\affiliation{Department of Applied Physics, University of Geneva, 1211 Geneva, Switzerland}
\affiliation{Constructor University Bremen, 28759 Bremen, Germany} 

\author{Julien Bertrand}
\affiliation{Department of Applied Physics, University of Geneva, 1211 Geneva, Switzerland}
\affiliation{Constructor University Bremen, 28759 Bremen, Germany}

\author{Mohsen Falamarzi Askarani}
\altaffiliation{Now at Xanadu Quantum Technologies Inc., Toronto, Ontario, M5G 2C8, Canada}
\affiliation{QuTech, Delft Technical University, 2628 CJ Delft, Netherlands}

\author{Maria Gieysztor}
\affiliation{QuTech, Delft Technical University, 2628 CJ Delft, Netherlands}
\affiliation{Institute of Physics, Faculty of Physics, Astronomy and Informatics, Nicolaus Copernicus University, Torun, Poland}

\author{Hridya Meppully Sasidharan}
\altaffiliation{Current address unknown.}
\affiliation{QuTech, Delft Technical University, 2628 CJ Delft, Netherlands}

\author{Joshua A. Slater}
\altaffiliation{Now at Q*Bird BV, 2628 XJ, Delft, The Netherlands.}
\affiliation{QuTech, Delft Technical University, 2628 CJ Delft, Netherlands}

\author{Aaron D. Marsh}
\altaffiliation{Now at Intel Corporation, Hillsboro, OR, USA}
\affiliation{Department of Physics, Montana State University, Bozeman, MT 59717-3840, USA}

\author{Philip J. T. Woodburn}
\altaffiliation{S2 Corporation, Bozeman, MT, USA}
\affiliation{Department of Physics, Montana State University, Bozeman, MT 59717-3840, USA}

\author{Charles W. Thiel}
\affiliation{Department of Physics, Montana State University, Bozeman, MT 59717-3840, USA}

\author{Rufus L. Cone}
\affiliation{Department of Physics, Montana State University, Bozeman, MT 59717-3840, USA}

\author{Sara Marzban}
\altaffiliation{Now at PiCard Systems BV, 6525EC Nijmegen, The Netherlands}
\affiliation{QuTech, Delft Technical University, 2628 CJ Delft, Netherlands}
%Toernooiveld 100, 

\author{Nir Alfasi}
\altaffiliation{Now at Quantum Machines, Tel Aviv, Israel}
\affiliation{QuTech, Delft Technical University, 2628 CJ Delft, Netherlands}

\author{Patrick Remy}
\affiliation{SIMH Consulting, Rue de Genève 18, 1225 Chêne-Bourg, Switzerland}

\author{Wolfgang Tittel}
 \email{Corresponding author: wolfgang.tittel@unige.ch}
\affiliation{Department of Applied Physics, University of Geneva, 1211 Geneva, Switzerland}
\affiliation{Constructor University Bremen, 28759 Bremen, Germany}

\date{\today}% It is always \today, today,

\begin{abstract}
We characterize spectroscopic and coherence properties of the 1451.37 nm excited-state zero-phonon line (ZPL) between the \textsuperscript{3}F\textsubscript{4}  and the \textsuperscript{3}H\textsubscript{4} manifolds of a thulium-doped yttrium aluminum perovskite (Tm$^{3+}$:YAlO$_3$) crystal at temperatures around 1.5 K. We measure the absorption spectrum between the\textsuperscript{3}H\textsubscript{6} - \textsuperscript{3}F\textsubscript{4}  and \textsuperscript{3}F\textsubscript{4} - \textsuperscript{3}H\textsubscript{4} manifolds, the inhomogeneous broadening of the  \textsuperscript{3}F\textsubscript{4} - \textsuperscript{3}H\textsubscript{4} (excited-state) ZPL, and the lifetimes of the higher-lying and lower-lying excited states. We also investigate level shifts caused by the quadratic Zeeman interaction as well as spectral hole-burning spectra with varying magnetic fields, providing insights into hyperfine interactions. Using again spectral holes but also optical free induction decays (FIDs), we assess optical coherence times, finding a maximum of $ 4.75 \pm 0.07$ $\mu s$ at B=2T and low ion concentration in the \textsuperscript{3}F\textsubscript{4} level. Our results---the first to demonstrate coherence of an excited-state transition in a rare-earth crystal---suggest the possibility of exploiting such transitions for quantum technology.
\end{abstract}

\maketitle

\noindent\emph{Introduction:}
Starting with the development of protocols that enable the storage and faithful recall of quantum states of light 20 years ago \cite{kraus2006quantum, alexander2006photon, nilsson2005solid}, trivalent rare-earth ions doped into cryogenically cooled inorganic crystals have played an increasingly important role in the development of quantum technology \cite{tittel2025quantum}. Today, ensemble-based quantum memories have reached a level of performance that allows early demonstrations of quantum repeater infrastructure \cite{sangouard2011quantum, Sinclair2014, liu2021heralded, Lago_Rivera2021, Ortu2022, Zhu2025}, and individual ions are being explored as single-photon sources \cite{Dibos2018,Zhong2018} and as qubits that allow quantum information processing \cite{Raha2020,Kindem2020,Gritsch2025}. At the heart of these applications is an important fundamental property -- highly coherent transitions between electronic and spin levels with coherence times that can exceed milliseconds \cite{Equall1994,bottger2009effects,Welinski2020,askarani2021long} and hours \cite{zhong2015optically}, respectively (see the supplemental material for details).

Surprisingly, while transitions between different (long-lived) excited states play an important role in solid-state lasers, e.g. the Nd-YAG laser \cite{geusic1964laser}, optical coherence at cryogenic temperatures of a few Kelvin has so far only been investigated for transitions starting at an electronic ground state. However, the long excited-state lifetimes that characterize the 4f levels of rare-earth ions allow, in principle, similarly long coherence times for excited-state transitions. In addition to being of fundamental interest, this would extend the use of rare-earth crystals to new wavelengths and novel applications. Here we investigate spectroscopic properties and optical coherence of the \textsuperscript{3}F\textsubscript{4} to \textsuperscript{3}H\textsubscript{4} excited-state transition in Tm\textsuperscript{3+}:YAlO\textsubscript{3}. 

\captionsetup[subfigure]{labelformat=empty}
\begin{figure*}[t] % * makes it span across two columns
    \centering
    % First subfigure
    \begin{subfigure}[t]{0.32\textwidth}
        \centering
        \caption{}\includegraphics[width=\linewidth,height=4.0cm,keepaspectratio]{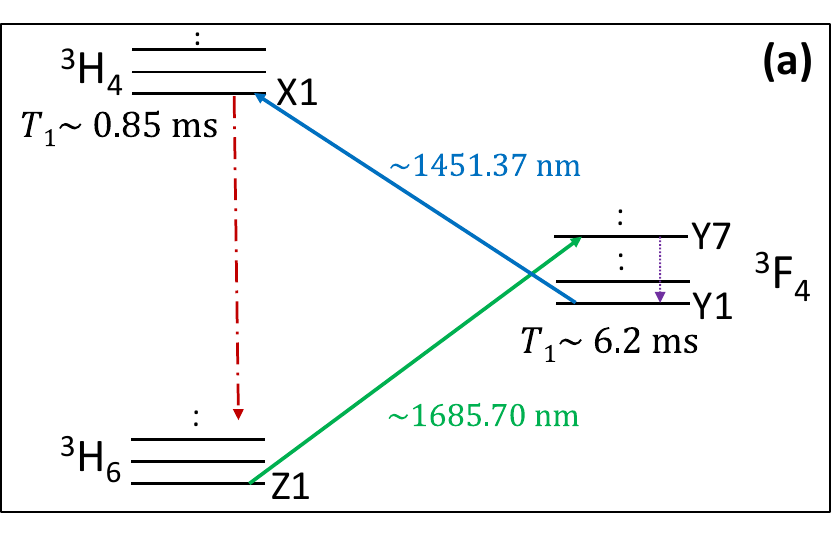}
        \label{fig:level_scheme}
    \end{subfigure}
    \hfill
    % Second subfigure
    \begin{subfigure}[t]{0.32\textwidth}
        \centering
        \caption{}
        \includegraphics[width=\linewidth]{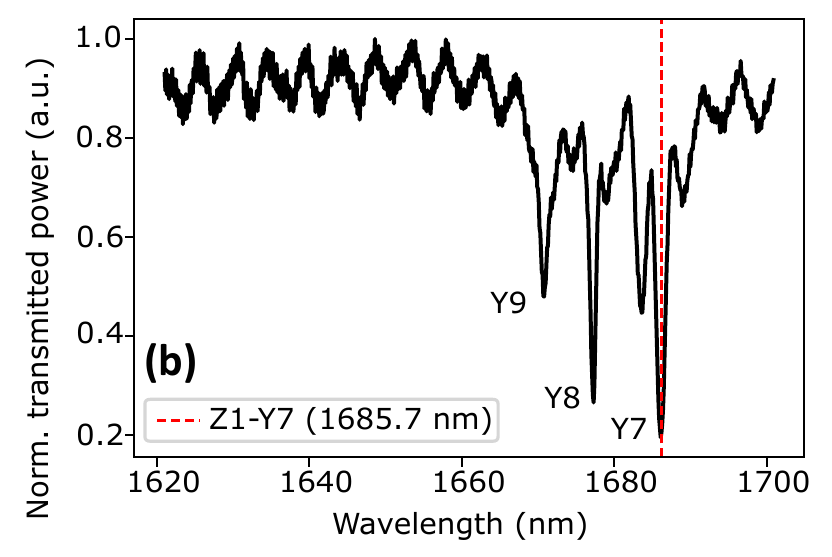}
        \label{fig:absorption_to_3F4}
    \end{subfigure}
    \hfill
    % Third subfigure
    \begin{subfigure}[t]{0.32\textwidth}
        \centering
        \caption{}
        \includegraphics[width=\linewidth]{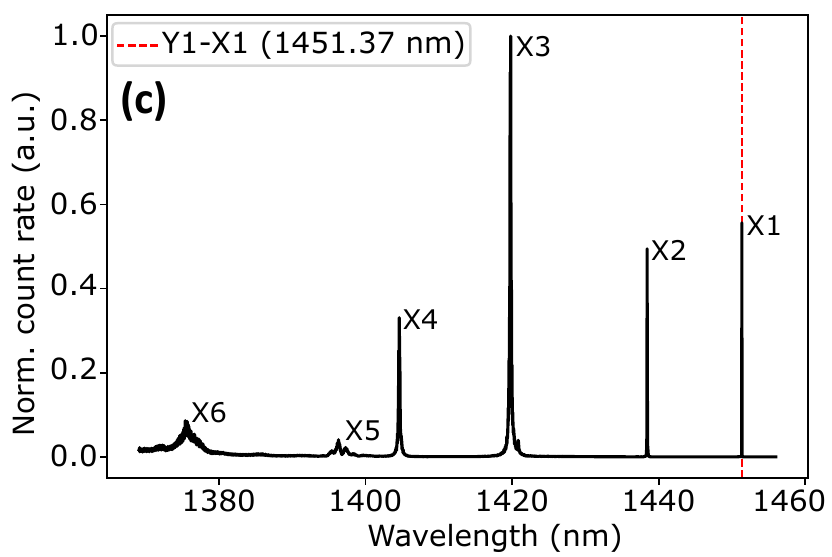}
        \label{fig:absorption_to_3H4}
    \end{subfigure}

    \caption{(a) Energy levels of relevant transitions of Tm\textsuperscript{3+} in YAlO$_3$. (b) Transmission spectrum measured using a broadband LED, showing absorption of the Z1 to \textsuperscript{3}F\textsubscript{4} transitions. (c) Photoluminescence rate at $\sim 800 \mathrm{nm}$ as a function of wavelength of a laser exciting the Y1 to \textsuperscript{3}H\textsubscript{4} transitions. Crystal field levels of the upper manifolds are identified in panels b) and c). }
    \label{fig:panel}
\end{figure*}

Several properties make Tm\textsuperscript{3+}:YAlO\textsubscript{3} a particularly interesting material to study. First, the transitions between its \textsuperscript{3}F\textsubscript{4} and the \textsuperscript{3}H\textsubscript{4} electronic manifolds comprise a so-called zero-phonon line (ZPL), i.e. a potentially coherent transition between their lowest Stark levels (or crystal field levels) (see figure \ref{fig:level_scheme} for a simplified level scheme that shows the ZPL between Y1 and X1).

Second, the wavelength of the \textsuperscript{3}F\textsubscript{4} to \textsuperscript{3}H\textsubscript{4} ZPL is around 1450 nm, where transmission loss in standard optical fiber is almost as small as in the telcom C-band, 0.25 dB/km and 0.2 dB/km, respectively. This makes Tm$^{3+}$:YAlO$_3$ a potential alternative to erbium-doped crystals, which feature a ground-state transition at around 1532 nm with optical coherence times up to 4.38 ms \cite{bottger2009effects}. Er-doped crystals have been exploited both for single-photon sources and quantum memory for light \cite{yu2023frequency, saglamyurek2011,miyazono2016coupling}.

And third, also shown in figure \ref{fig:level_scheme}, Tm\textsuperscript{3+}:YAlO\textsubscript{3} features three ZPLs---two ground-state transitions and one excited-state transition---that are arranged in a ring. This may allow the development and implementation of novel protocols that require more than a single line, e.g. the creation of energy-time entangled photon pairs using two cascading transitions (transitions X1 - Y1 and Y1 - Z1 in Fig.\ref{fig:level_scheme}) as originally proposed in \cite{franson1989bell}, and quantum computing using one transition for optical state read-out (e.g. transition Z1 - X1) and another transition (e.g. transition Z1 - Y1) for dipole-blockade-based controlled 2-qubit gates, similar to the case of trapped ions \cite{Myerson2008}.

The energy-level structure of Tm\textsuperscript{3+}:YAlO\textsubscript{3} has been characterized before \cite{antonov1973spectral, o1976crystal,korner2018spectroscopic,guillemot2022excited}. However, to the best of our knowledge, our investigation is the first to study the excited-state transition of Tm\textsuperscript{3+}:YAlO\textsubscript{3}--- more precisely of any rare-earth crystal---at cryogenic temperatures where coherence properties can be assessed.

Our paper is organized as follows.  First, we measure the absorption spectrum between the  \textsuperscript{3}H\textsubscript{6} and \textsuperscript{3}F\textsubscript{4} manifolds, and we describe how we populate the Y1 level. We then extend the absorption measurements to transitions between the \textsuperscript{3}F\textsubscript{4} and \textsuperscript{3}H\textsubscript{4} manifolds, and we characterize the inhomogeneous broadening of the ZPL at 1451.37 nm as well as the lifetimes of the Y1 and X1 levels. Moving on to magnetic interactions, we study the quadratic Zeeman shift affecting the $\sim$1451.3 nm transition, and we study level splittings by means of spectral hole burning. We also assess optical coherence as a function of magnetic field, wait time between excitation and read-out, and ion concentration in the Y1 level using again spectral holes as well as free induction decays. To round off the paper, we discuss next steps towards applications of Tm\textsuperscript{3+}:YAlO\textsubscript{3} in quantum information processing.
 
\vspace{0.2cm}

\noindent\emph{Experimental details:}
YAlO$_3$ features an orthorombic unit cell with lattice constants a=5.179 \AA, b=5.329 \AA, c = 7.370 \AA (Pbnm space group). Tm$^{3+}$ substitutes Y$^{3+}$ ions in four structurally inequivalent positions with low point-group symmetry C$_{1h}$ \cite{o1976crystal}. Doped Tm$^{3+}$ ions are related pairwise by inversion symmetry, resulting in two pairs of sites that become magnetically equivalent for magnetic fields directed within the ac and bc planes \cite{asatryan2002epr}, where a, b and c denote the principal crystal axes.

Measurements are carried out using a a=12.0 mm $\times$ b=2.2 mm $\times$ c= 2.0 mm large 0.1\% crystal (from Scientific Material Corp.). The crystal is glued to a customized Cu stage and further secured and thermalized by a Cu lid with spring-tensioned screws. It is cooled to around 1.5 K using a pulse-tube cooler supplemented by an adiabatic demagnitization stage. A superconducting solenoid allows one to apply a homogeneous magnetic field between 0 and 2 T approximately along the b-axis of the crystal.

The laser light is sent through the crystal along the a-axis using a single-mode (SM) pigtailed ferrule and a gradient-index (GRIN) lens. The ferrule is attached to an attocube nanopositioner, which allows adjusting the distance between the ferrule and the GRIN lens to collimate the laser beam to a diameter of around 0.3 mm. Behind the crystal, a mirror reflects the light back through the crystal and into the fiber. A tunable continuous-wave (CW) laser (Toptica DL pro) or an SLED (Denselight Semiconductors DL-BZ1) emitting at around 1680 nm wavelength  is used for optical pumping, and another CW laser (Santec TSL-550) drives the excited-state transition at around 1450 nm wavelength. One fiber-based acousto-optic modulator (AOM) per laser allows creating short pulses, and a phase modulator (PM) is used to modify the spectrum of the $\sim$ 1450 nm laser light. 

Depending on the experiment, we can choose between different detectors. To detect photoluminescence at 795 nm wavelength caused by atomic decay from the $^3$H$_4$ to the $^3$H$_6$ manifold, a single-photon counting module (SPCM) based on a silicon-avalanche photodiode is used (from Perkin Elmer). In this case, we also employ free-space spectral filters (a low-pass filter composed of multiple EO2-coated mirrors and a DMSP950 dichroic mirror, all from Thorlabs) that reject light at 1680 nm and 1450 nm. For spectral hole burning measurements, the 1450 nm laser light is directed to an InGaAs photoreceiver with 10 MHz bandwidth (New Focus 2053). Finally, to characterize free induction decays (FIDs), a higher-bandwidth photoreceiver (New Focus 1811) with a rise and fall time of 3 ns is used. It is preceded by a gating AOM that blocks irrelevant pulses. For a schematic of the optical setup, see the supplemental material.

\vspace{0.2cm}

\begin{figure}[h]
    \includegraphics[width=0.8\columnwidth]{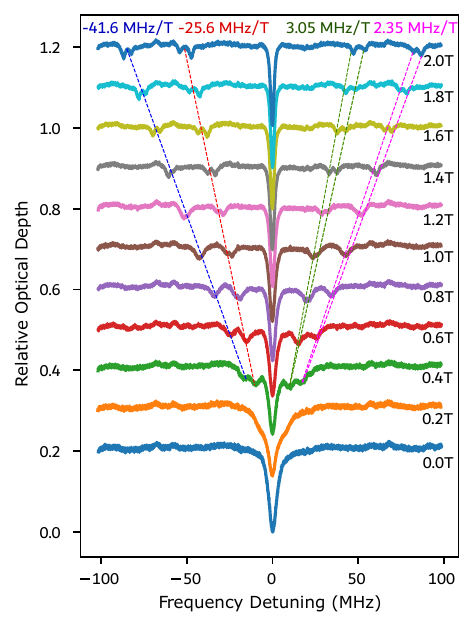}
    \caption{Spectral hole burning spectra under varying magnetic fields. Traces measured with different fields are offset by an amount that is proportional to the field, and splitting rates are indicated.}
    \label{fig:shb}
\end{figure}

\noindent\emph{Optical pumping:}
At 1.5 Kelvin, the Tm\textsuperscript{3+} ions occupy the lowest crystal field level Z1 of the \textsuperscript{3}H\textsubscript{6} ground-state multiplet. Since we are interested in the Y1 to X1 transition between the \textsuperscript{3}F\textsubscript{4} and the \textsuperscript{3}H\textsubscript{4} manifolds, our first goal is to populate the Y1 level. Towards this end, we can pump ions into any crystal field level within the \textsuperscript{3}F\textsubscript{4} manifold. Since intra-manifold relaxation between different levels takes place on the nanosecond time scale \cite{thiel2011rare}, all ions in levels Y2 to Y9 will rapidly decay into Y1. This approach to pumping is promising since it allows one to interact continuously with an empty upper atomic level. In addition, one can choose the strongest absorption line. An alternative possibility is described in the supplemental materials.

\vspace{0.2cm}

\noindent\emph{Absorption spectra:}
We study the absorption lines between Z1 and the \textsuperscript{3}F\textsubscript{4} manifold by measuring the wavelength-dependent transmission of the broadband SLED  using an optical spectrum analyzer (Anritsu MS9710C), see figure \ref{fig:absorption_to_3F4} (see \cite{antonov1973spectral} for the association of absorption lines with energy levels). After optimization of polarization using a fiber-optic polarization controller, we find that the Z1 to Y7 transition at 1685.7 nm features the strongest absorption. It is used in the following to populate the Y1 level. 

Next, we measure the absorption spectrum of the Y1 to \textsuperscript{3}H\textsubscript{4} transitions, still at B=0T. After populating Y1 using the 1686 nm laser, we expose the crystal to laser light whose wavelength we scan between 1370 nm and 1456 nm in steps of 0.01 nm. The detection of photoluminescence (PL) around 800 nm indicates resonances and, at the same time, allows optimizing the laser polarization and the initial pumping step. Our measurement results are shown in figure \ref{fig:absorption_to_3H4}. In particular, we find a well isolated line at 1451.37 nm -- the ZPL between Y1 and X1 \cite{o1976crystal}. A subsequent laser scan around this wavelength with improved resolution of 0.001 nm reveals a linewidth of 1.29 GHz. This is much narrower than any of the nm-wide ground-state transitions shown in figure \ref{fig:absorption_to_3F4}. We furthermore measure an optical depth of up to 1.12 on line center, which we expect to increase with optimized pumping (see the supplemental material). 

Small measurment variations allow establishing the radiative lifetimes of the Y1 and the X1 levels. For the former, we monitor the time-resolved decay of the optical depth of the Y1 to X1 transition after switching off the 1686 nm laser. For the latter, we analyze the decay of the 800 nm PL rate after pulsed excitation of the X1 level. We find lifetimes of 6.21 $\pm$ 0.01 $\mathrm{ms}$ and 845.4 $\pm$ 2.4 $\mathrm{\mu s}$  for the \textsuperscript{3}F\textsubscript{4} and the \textsuperscript{3}H\textsubscript{4} levels, respectively, which is consistent with literature values \cite{guillemot2019continuous} and an alternative approach detailed in the supplemental material.

\vspace{0.2cm}

\noindent\emph{Level shifts and splitting:}
The magnetic field-dependent change of the Tm energy levels is described by the Hamiltonian
\begin{equation}
H=\mathbf{B}\cdot(-g_n\beta_n\opone -2g\mu_BA\,\mathbf{\Lambda})\cdot\mathbf{I}-g^2\mu_B^2\mathbf{B}\cdot \mathbf{\Lambda}\cdot\mathbf{B}
\end{equation}
where the first part describes level splitting due to the enhanced nuclear Zeeman effect (the combination of the nuclear Zeeman interaction and the second-order coupling between the hyperfine interaction and the electronic Zeeman effect) and the last part describes level shifts caused by the quadratic Zeeman interaction. Here, $g_n$ is the nuclear g-factor of Tm, $\beta_n$ the nuclear magneton, $g$ the Landé g-factor for each level, $\mu_B$ the Bohr magneton, $A$ the hyperfine constant, $\mathbf{I}$ the ion's nuclear spin, $\mathbf{B}$ the applied magnetic fild, and $\Lambda$ the hyperfine tensor \cite{davidson2021measurement}.

First, to quantify the quadratic Zeeman effect, we measure the shift of the Y1 to X1 transition as a function of applied magnetic field, yielding $\Delta\nu= -101 \pm 3$ MHz/T$^2$ (see  supplemental material). 

Second, to assess the magnetic field-dependent energy-level splitting of the thulium ions as a consequence of the enhanced nuclear Zeeman effect, we perform a series of spectral hole burning experiments. After populating Y1, we expose the Tm ions during a \emph{burn time} of 300 $\mu s$ to narrow-bandwidth 1451 nm laser light. After 10 $\mu s$, we start scanning the 1451 nm laser over a 200 MHz wide spectral interval that is centered on the burn frequency and we monitor the transmitted optical power. We define the \emph{wait time} as the time between the end of the burn pulse and the moment when the scanning laser features the same frequency as the burn pulse, i.e. the moment when it scans over the previously created spectral hole, here 85 $\mu s$. For these measurements, we kept the density of the ions in the \textsuperscript{3}F\textsubscript{4} level constant and we burned the spectral hole at the center of the Y1$\leftrightarrow$X1 (1451 nm) line -- the optical depth was around 0.4. Figure \ref{fig:shb} shows the resulting hole-burning spectra for magnetic fields between 0 and 2T, consisting of a central hole and two clusters of side holes on either side. We find that the width of the central hole decreases with increasing magnetic field, that the center of each cluster of side holes shifts linearly at rates of $\pm$ 25.6 MHz/T and  $\pm$ 41.6 MHz/T away from the center, and that the two holes within each cluster also split as a function of the field at a rate of 2-3 MHz/T (We attribute this sub-structure to a slight misalignment of the magnetic field with respect to the crystal b-axis, in which case the two pairs of Tm-sites feature different level splittings.). We also note the absence of anti-holes, i.e. narrow spectral regions with increased absorption.

To explain the origin of the clusters, we take into account that Tm has a nuclear spin $I = 1/2$, and that the hyperfine coupling between the nuclear spin and the electronic states splits each crystal field level into a pair of hyperfine levels \cite{sinclair2021optical}. As we further illustrate in the supplemental material, narrow-bandwidth optical pumping therefore drives four different transitions within the ensemble of inhomogeneously broadened thulium ions -- one between each combination of a ground state and an excited state. This results in decreased absorption -- a central hole and four side holes -- at five different frequencies. Note that the sum of the burn and wait time is sufficiently small compared to the lifetime of the two magnetic X1 levels to exclude significant decay of atoms from X1 back into Y1 (in addition, the branching ratio for this pathway is rather small, and most decay leads back to the ground state). This explains the absence of anti-holes.

Next, we perform time-resolved spectral hole burning. We find that the two ``inner" clusters (with splitting rate of $\pm$25.6 MHz/T) decay on a scale comparable to the radiative lifetime of X1 (0.85 ms). As explained in the supplemental material, this suggests that these clusters are caused by the splitting of the two hyperfine levels within the Y1 doublet. The other pairs of clusters, with splitting rate of $\pm$41.6 MHz/T, decay with a much longer lifetime, comparable to that of the central hole. We conjecture that they are caused by the splitting of the two hyperfine levels within the X1 doublet.  

\begin{figure*}[ht]
    \centering
    % Panel (a)
    \begin{subfigure}{0.32\textwidth}
        \centering
        \caption{}
        \includegraphics[width=\linewidth]{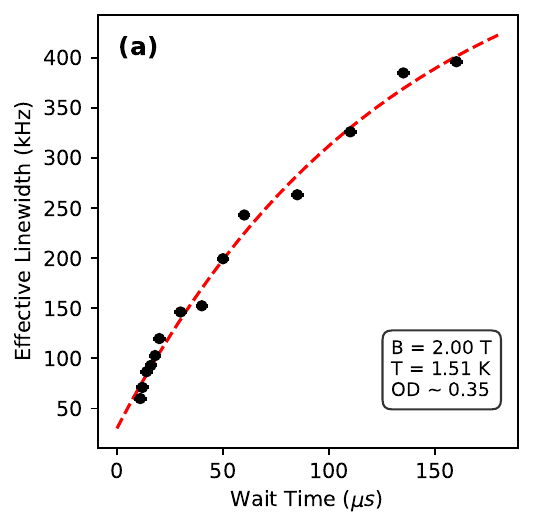}
        \label{fig:time_resolved_SHB_for_SD}
    \end{subfigure}
    \hfill
    % Panel (b)
    \begin{subfigure}{0.32\textwidth}
        \centering
        \caption{}
        \includegraphics[width=\linewidth]{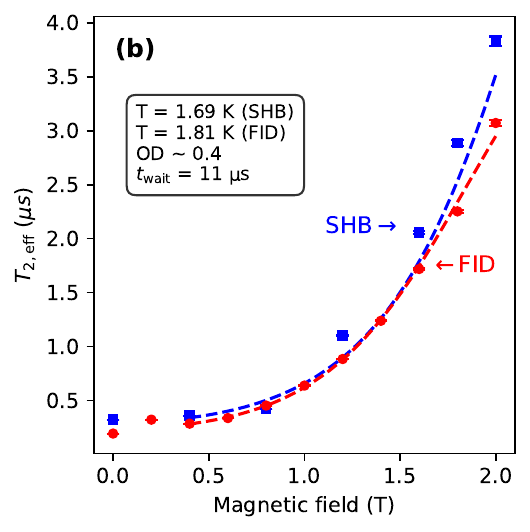}
        \label{fig:coherence}
    \end{subfigure}
    \hfill
    % Panel (c)
    \begin{subfigure}{0.32\textwidth}
        \centering
        \caption{}
        \includegraphics[width=\linewidth]{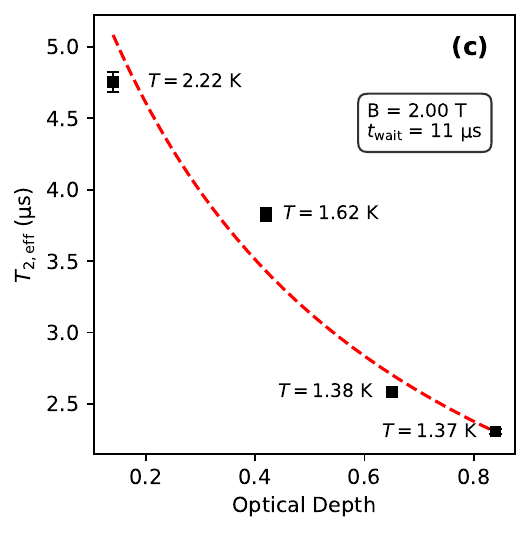}
        \label{fig:OD_dependence}
    \end{subfigure}
    \caption{(a) Effective linewidths measured using spectral holes with varying wait time and B= 2T. The red dashed line is a fit based on Eq. \ref{eq:gamma_eff}.
    (b) Effective optical coherence times for different magnetic fields obtained by means of spectral hole burning (blue squares, $t_{\text{wait}}$=11 $\mu s$) and free induction decay (red circles). The dashed lines are fits based on a spectral diffusion model \cite{bottger2006optical}.
    (c) Effective optical coherence times measured using SHB as a function of OD. The temperature varied significantly between measurements. 
    The dashed line is a fit based on a model considering instantaneous spectral diffusion \cite{thiel2014measuring}. All data points in panels a-c are corrected with respect to the laser linewidth of 170 kHz, and error bars are based on an uncertainty of that linewidth of $\pm$ 2 kHz.}
\end{figure*}

\vspace{0.2cm}
\noindent\emph{Optical coherence:}
In addition to providing insight into hyperfine interactions, the hole-burning spectra---more precisely the width of the central holes---also allow one to derive an upper bound for the homogeneous linewidth $\gamma_{\text{hom}}$ and hence a lower bound for the optical coherence time $T_2= 1/(\pi \gamma_{\text{hom}})$ of the excited-state transition. Indeed, one can show that $\gamma_{\text{hole}} = \gamma_{\text{hom}}\left( 1 + \sqrt{1 + \chi^2 T_1T_2} \right) + \gamma_{\text{laser}}$ \cite{de1980photophysical}, where $\gamma_{\text{hole}}$ is the full-width-at-half-maximum (FWHM) of the hole, $\chi$ is the on-resonance Rabi frequency, and $T_1$ is the radiative life time. Furthemore, $\gamma_{\text{laser}}$ describes the linewidth of the laser used for hole burning, which we measured by delay self-homodyne interferometric detection\cite{bai2021narrow} to be $171.1 \pm 2.0$ kHz. When the power of the burn pulse is sufficiently small (and $\chi^2 \approx 0$), this equation can be simplified to $\gamma_{\text{hole}} \approx 2\gamma_{\text{hom}} + \gamma_{\text{laser}}$. 

First, to characterize the effect of spectral diffusion, we analyze the evolution of the measured (effective) homogeneous linewidth as a function of wait time $t_{\mathrm{wait}}$. Following \cite{bottger2006optical}, it can be described as

\begin{equation}
    \gamma_{\text{eff}} = (\gamma_{\text{hole}}-\gamma_{\text{laser}})/2 = \gamma_0 + \frac{\gamma_{\text{SD}}}{2} [1 - e^{-Rt_{\mathrm{wait}}}],
    \label{eq:gamma_eff}
\end{equation}
where $\gamma_{0}$ is the intrinsic linewidth (no spectral diffusion), and $\gamma_{\text{SD}}$ and $R$ describe the maximum broadening due to spectral diffusion and the rate with which the maximum is reached, respectively. The latter equals the average spin-flip rate of the spin bath around the probed ions.

Figure \ref{fig:time_resolved_SHB_for_SD} shows $\gamma_{\text{eff}}$ as a function of  $t_{\mathrm{wait}}$, varied between $11 \mu s$ and $161 \mu s$ at B = 2T. The data are well described by Eq.\ref{eq:gamma_eff}, and show the presence of significant spectral diffusion on the timescale of our measurements. The fit gives an average bath spin-flip rate $R = 7.82 \pm 1.75 \text{kHz}$, which corresponds to a characteristic spectral diffusion timescale of $1/R \approx 128 \mu s$, and a maximum linewidth $\gamma_{\text{eff}}^{\text{max}}=550.9 \pm 60.3$ kHz (for $\gamma_0=30.0\pm 9.7$ kHz and $\gamma_{\text{SD}}=1041.7\pm 119.1$ kHz).

Figure \ref{fig:coherence} depicts $T_{\text{2,eff}}$ for a fixed wait time of $t_{\text{wait}}=11\mu s$ and different magnetic fields, where  $T_{\text{2,eff}} = 1/(\pi \gamma_{\text{eff}})$ is the effective optical coherence time.
\vspace{0.2cm}

Another way to assess the optical coherence time is via the free induction decay (FID) \cite{brewer1972optical}. In this case, ions are excited using a pulsed laser (pulse duration of 20 $\mu s$) with a bandwidth that is smaller than the homogeneous linewidth of the transition to be studied. The thereby created atomic polarization gives rise to the emission of light, which decreases exponentially after the end of the laser pulse due to atomic dephasing. 

As shown in the supplemental material, $\tau_\text{FID} = 1/\left[2\pi (2\gamma_\text{eff}+\gamma_\text{laser})\right]$, where $\tau_{\text{FID}}$ is the 1/e decay of the intensity of the FID signal. After solving for $\gamma_{\text{eff}}$, this leads to 
\begin{equation}
T_{\text{2,eff}} =\frac{4}{1/\tau_{\text{FID}}-2\pi\gamma_{\text{laser}}}.
\label{eq:T2_eff_FID}
\end{equation}

Figure \ref{fig:coherence} depicts  $T_{\text{2,eff}}$ as a function of magnetic field, along with the results obtained using SHB, as described above. We note the good match between the two data sets. The figure also shows fits of a spectral diffusion model that considers magnetic noise from nearby spin flip-flops and phonon driven spin flips \cite{bottger2006optical} to both sets. We find that the effect of spin flips can be ignored in the assessed parameter regime. See the supplemental material for more information.

Further studies are required to identify the source of spin flip-flops. The nuclear spins of neighboring lattice ions (most likely Al\textsuperscript{3+} \cite{bai1991spin, yano1992stimulated}) are possible contributors, even though their impact seems too small to explain our observations. But note that the energy difference between the Z1 and Z2 levels in the \textsuperscript{3}H\textsubscript{6} manifold of Tm\textsuperscript{3+} is very small ($\sim90\text{GHz}$). These levels may therefore form non-Kramers quasi-doublets that exhibit a strong magnetic moment, the so-called Van Vleck paramagnetism \cite{van1932theory, takikawa2010van}, which may also contribute to magnetic noise. See the supplemental material for more details.

Finally, we  vary the ion density in Y1 by changing the optical power during pumping. Since we continue burning spectral holes on line center of the 1451 nm transition, we can parametrize this change in terms of the optical depth of the Y1$\leftrightarrow$X1 transition. As shown in figure \ref{fig:coherence} for B=2T and $t_{\text{wait}}$=11$\mu s $, $T_{\text{2,eff}}$ increases to $4.75 \pm 0.07 \mu s$ as the optical depth is reduced from 0.84 to 0.14. We attribute this dependence to flip-flops between spins of neighboring Tm ions in the Y1 level, or, more likely, to instantaneous spectral diffusion (ISD) \cite{thiel2014measuring} caused by exciting Tm ions to the X1 level. 
Note that the temperature varied significantly during these measurements. But since we observed an increase of coherence time with increasing temperature and not the opposite, these variations do not allow explaining our results. 

\vspace{0.2cm}
\noindent\emph{Conclusion:}
We have investigated an excited-state transition in Tm\textsuperscript{3+}:YAlO$_3$ at cryogenic temperatures and varying magnetic fields. We observed a quadratic Zeeman shift of the transition wavelength as well as level spittings caused by the enhanced nuclear Zeeman interaction. In addition, we measured for the first time coherence of such a transition in any rare-earth ion doped crystal. While our findings clearly show that optical coherence of the excited state transition will continue to increase as the magnetic field is increased beyond 2T, and $t_{\text{wait}}$ and the optical depth are further decreased, more measurements are needed to better understand and mitigate the origin of significant spectral diffusion.  

Several avenues can be pursued. First, spin flips and flip-flops are suppressed at lower temperature and a longer optical coherence time can thus be expected by further cooling the crystal. For instance, the spectral diffusion model used to fit the data in Figure 3b predicts $T_2=10.6 \mu s$ for T$<$0.8K, B=2T, OD=0.4 and $t_{\text{wait}}=11\mu s$, see supplemental material. Second, reducing either the Tm doping concentration, the concentration of Tm-ions in the Y1 level, and/or the optical power used to burn spectral holes is expected to lead to a further increase of $T_2$, with values of several ten $\mu$s appearing realistic to reach. Third, the magnetic field can be increased up to a point where spin flips become the primary limitation. This should also reduce the impact of the Van Vleck paramagnetism as the energy gap between the lowest Stark levels in the \textsuperscript{3}H\textsubscript{6} manifold will increase and flip-flops of pseudo-spins therefore become less likely. And fourth, changing both the magnitude and the direction of the magnetic field may lead to spectrally stable optical clock transitions, which have been reported before for several rare-earth crystals \cite{Ortu2018,huang2023investigation}, including for a Tm-doped crystal \cite{davidson2021measurement}. Our finding of a quadratic Zeeman shift is an important first step towards this end.

Overall, our results demonstrate that excited-state transitions in REI-doped crystals may be useful for applications in quantum information processing. In particular, the 1450 nm line in Tm may enable to creation of true single-photon sources and quantum memory within a telecom band -- a potential alternative to erbium and possibly enabling the creation of quantum networks coexisting with classical communication at 1550 nm over the same fiber infrastructure \cite{valivarthi2019}.
 
\vspace{0.2cm}
\noindent\emph{Acknowledgements---}The authors thank Javier Carrasco \'Avila, Archi Gupta and Deeksha Gupta for discussions. WT acknowledges funding through the Netherlands Organization for Scientific Research, the European Union’s Horizon 2020 Research and Innovation Program under Grant Agreement No. 820445 and Project Name Quantum Internet Alliance, and the Swiss State Secretariat for Education, Research and Innovation (SERI) under Contract Number UeM019-3. Work at MSU was supported in part by the Air Force Research Laboratory under award number FA8750-23-2-0500.

\vspace{0.2cm}
\noindent\emph{Data availability---}
The data that support the findings of this article are openly available \cite{li_2026_20024931}.

\bibliography{references}
\end{document}

% \preprint{APS/123-QED}

\title{Spectroscopy and Coherence of an Excited-State Transition in Tm\textsuperscript{3+}\!:\,YALO\textsubscript{3} at Telecommunication Wavelength --  Supplemental Material}% 

\author{Luozhen Li}
 \altaffiliation{These authors contributed equally to this work.}%Lines break automatically or can be forced with \\
\affiliation{Department of Applied Physics, University of Geneva, 1211 Geneva, Switzerland}
\affiliation{Constructor University Bremen, 28759 Bremen, Germany} 

\author{Akshay Babu Karyath}
 \altaffiliation{These authors contributed equally to this work.}
\affiliation{Department of Applied Physics, University of Geneva, 1211 Geneva, Switzerland}
\affiliation{Constructor University Bremen, 28759 Bremen, Germany} 

\author{Julien Bertrand}
\affiliation{Department of Applied Physics, University of Geneva, 1211 Geneva, Switzerland}
\affiliation{Constructor University Bremen, 28759 Bremen, Germany}

\author{Mohsen Falamarzi Askarani}
\altaffiliation{Now at Xanadu Quantum Technologies Inc., Toronto, Ontario, M5G 2C8, Canada}
\affiliation{QuTech, Delft Technical University, 2628 CJ Delft, Netherlands}

\author{Maria Gieysztor}
\affiliation{QuTech, Delft Technical University, 2628 CJ Delft, Netherlands}
\affiliation{Institute of Physics, Faculty of Physics, Astronomy and Informatics, Nicolaus Copernicus University, Torun, Poland}

\author{Hridya Meppully Sasidharan}
\altaffiliation{Current address unknown.}
\affiliation{QuTech, Delft Technical University, 2628 CJ Delft, Netherlands}

\author{Joshua A. Slater}
\altaffiliation{Now at Q*Bird BV, 2628 XJ, Delft, The Netherlands.}
\affiliation{QuTech, Delft Technical University, 2628 CJ Delft, Netherlands}

\author{Aaron D. Marsh}
\altaffiliation{Now at Intel Corporation, Hillsboro, OR, USA}
\affiliation{Department of Physics, Montana State University, Bozeman, MT 59717-3840, USA}

\author{Philip J. T. Woodburn}
\altaffiliation{S2 Corporation, Bozeman, MT, USA}
\affiliation{Department of Physics, Montana State University, Bozeman, MT 59717-3840, USA}

\author{Charles W. Thiel}
\affiliation{Department of Physics, Montana State University, Bozeman, MT 59717-3840, USA}

\author{Rufus L. Cone}
\affiliation{Department of Physics, Montana State University, Bozeman, MT 59717-3840, USA}

\author{Sara Marzban}
\altaffiliation{Now at PiCard Systems BV, 6525EC Nijmegen, The Netherlands}
\affiliation{QuTech, Delft Technical University, 2628 CJ Delft, Netherlands}
%Toernooiveld 100, 

\author{Nir Alfasi}
\altaffiliation{Now at Quantum Machines, Tel Aviv, Israel}
\affiliation{QuTech, Delft Technical University, 2628 CJ Delft, Netherlands}

\author{Patrick Remy}
\affiliation{SIMH Consulting, Rue de Genève 18, 1225 Chêne-Bourg, Switzerland}

\author{Wolfgang Tittel}
 \email{Corresponding author: wolfgang.tittel@unige.ch}
\affiliation{Department of Applied Physics, University of Geneva, 1211 Geneva, Switzerland}
\affiliation{Constructor University Bremen, 28759 Bremen, Germany}

\date{\today}% It is always \today, today,

\maketitle
\newpage

\tableofcontents

\section{Coherence times of rare-earth-doped crystals}
To provide additional context, table \ref{tab:T2} lists relevant coherence times of various transitions of rare-earth crystals.

\begin{table}[h]
    \centering
    \begin{tabular}{c|c|c|c|c|c}
        Crystal & Transition type & Transition & wavelength or frequency  & T$_2$ & Reference\\
        
        \hline\hline
        
        Eu:Y$_2$SiO$_5$ & ground-state (spin) & $^7F_0$ & 12.5 MHz & 18h & \cite{wang2025nuclear}\\

        Pr:Y$_2$SiO$_5$ & ground-state (spin) & $^3H_4$ & 10 MHz & 1 min & \cite{heinze2014coherence}\\
        
        Er:Y$_2$SiO$_5$ & ground-state (spin) & $^4I_{15/2}$ & 995 MHz  & 1.3 s &\cite{ranvcic2018coherence}\\
        
         % Eu:Y$_2$SiO$_5$ & ground-state (spin) & $^7F_0$ & 12.5 MHz & 18h & \cite{zhong2015optically}\\
         
         % Yb:Y$_2$SiO$_5$  & ground state (spin) & - & 2.372 GHz & 6 ms & \cite{alexander2022coherent}\\
         
         Yb:YVO$_4$  & ground state (spin) & $^2F_{7/2}$ & 2.4 GHz & 6.6 ms & \cite{Kindem2018}\\
         
         \hline
        
         Er:Y$_2$SiO$_5$  & excited state (spin) & $^4I_{13/2}$ & 896.7 MHz & 1.48 ms & \cite{Rakonjac2020}\\

         Er:Y$_2$SiO$_5$  & excited state (spin) & $^4I_{13/2}$ & 400 MHz & 1.6 $\mu$s & \cite{welinski2019electron}\\
         
         \hline
         Er:Y$_2$SiO$_5$ & ground-state (optical) & $^4I_{15/2}~\leftrightarrow~^4I_{13/2}$ & 1536 nm & 4.38 ms & \cite{bottger2009effects}\\

         Eu:Y$_2$SiO$_5$ & ground-state (optical) & $^7F_0~\leftrightarrow~^5D_0$ & 580 nm & 2.6 ms & \cite{Equall1994}\\
         
         Tm:Y$_3$Ga$_5$O$_{12}$ & ground-state (optical) & $^3H_6~\leftrightarrow~^3H_4$ & 795 nm & 1.1 ms & \cite{thiel2014optical}\\ 
         
        Yb:Y$_2$SiO$_5$ & ground-state (optical) & $^2F_{7/2}~\leftrightarrow~^2F_{5/2}$ & 980 nm & 0.8 ms & \cite{Welinski2020}\\

        Pr:Y$_2$SiO$_5$ & ground-state (optical) & $^3H_4~\leftrightarrow~^1D_2$ & 607 nm & 377 $\mu$s & \cite{equall1995homogeneous}\\
        
        Er:LiNbO$_3$ & ground-state (optical) & $^4I_{15/2}~\leftrightarrow~^4I_{13/2}$ & 1532 nm & 180 $\mu$s & \cite{wang2022er}\\
        
        \hline
        
        Tm:YAlO$_3$ & excited-state (optical) & $^3F_4 \leftrightarrow ^3H_4$ & 1451 nm & 4.75 $\mu$s & this work\\
    \end{tabular}
    \caption{Coherence times reported for different rare-earth doped crystals. In case of spin transitions, only the electronic manifold is indicated.}
    \label{tab:T2}
\end{table}

\section{Experimental schematic}
\begin{figure}[h]
    \centering
    \includegraphics[width=1\linewidth]{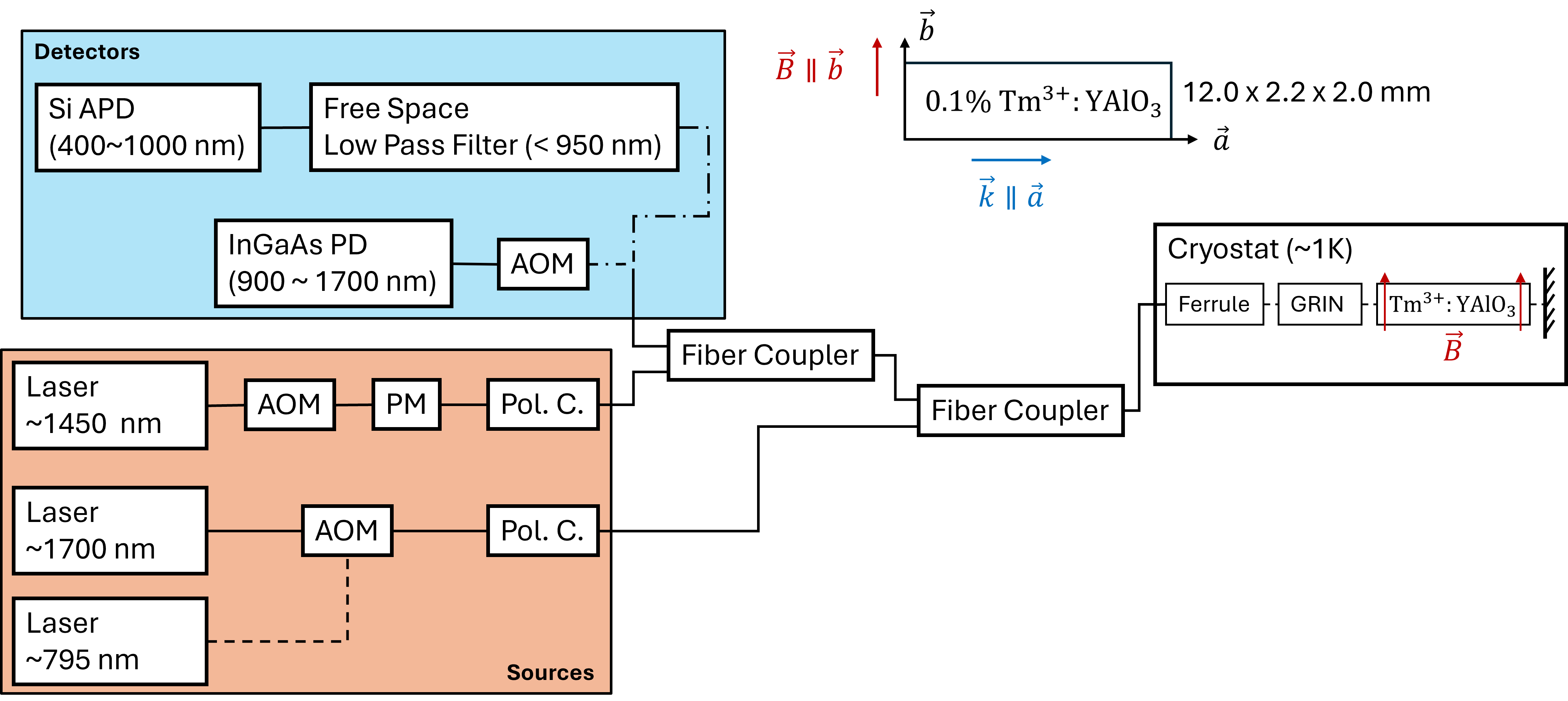}
    \caption{Schematic of experimental setup. AOM: acousto-optic modulator; PM: phase modulator; Pol. C.: fiber polarization controller; GRIN: gradient-index lens; PD: photodiode; Si APD: silicon-avalanche photodiode. Light propagates along the a-axis of the crystal. The magnetic field is applied along the b-axis of the crystal.}
    \label{fig:experimental setup}
\end{figure}

A generic schematic of the experimental setups is shown in Figure \ref{fig:experimental setup}.

\section{Populating Y1 via spontaneous decay from $^3$H$_4$}

There are two different ways of populating the Y1 level. The more efficient way is described in the main text. Here we detail our work based on an alternative approach in which we first pump population into the $^3$H$_4$ manifold using a 795 nm wavelength laser, and then wait for spontaneous decay to populate the Y1 level within the $^3$F$_4$ manifold. 

\subsection{Absorption spectra}
\begin{figure}[h]
    \centering
    \includegraphics[width=0.95\linewidth]{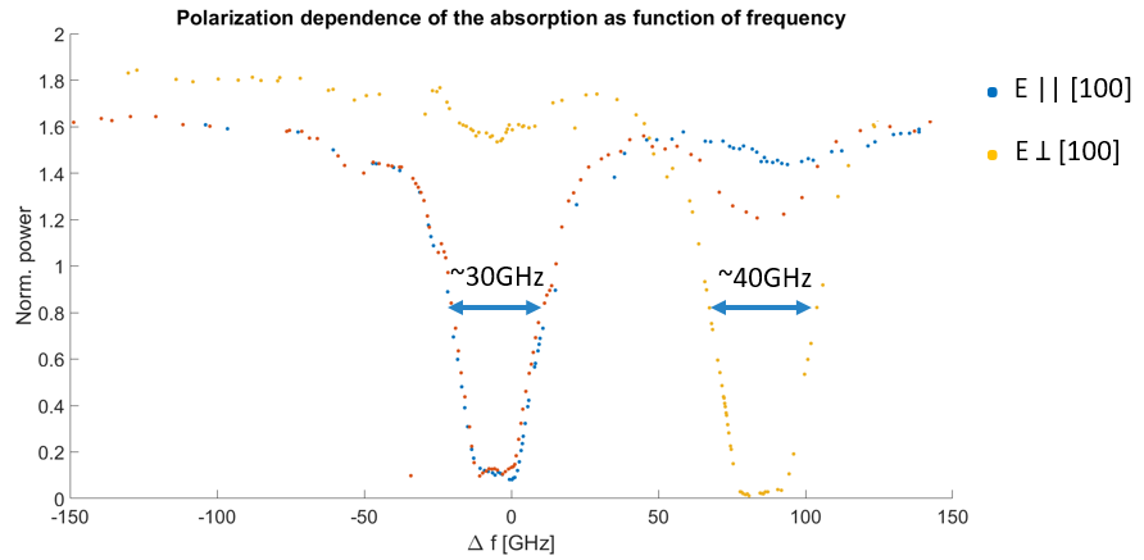}
    \caption{Two orthogonally polarized zero-photon lines at around 795 nm wavelength between Z1 and X1.}
    \label{fig:795nm_absorption}
\end{figure}

To optimize optical pumping, we again start by measuring absorption spectra, now including transitions at around 795 nm wavelength and between Z1 and the $^3$H$_4$ manifold. As we see in Figure \ref{fig:795nm_absorption}, we find two closely spaced absorption lines for orthogonally polarized light at $\sim$795.1\,nm and $\sim$795.3\,nm wavelength. Both are zero-phonon lines for different classes of centers. Given the larger optical depth, we decided to work with the line at 795.1 nm.

\subsection{Experimental procedure}
To assess the possibility of populating the Y1 energy level (the lowest crystal field level within the $^3$F$_4$ manifold), we employ a pump–wait–read sequence: A 795\,nm pump of duration $\tau_{\rm pump}$ excites the $^3$H$_4$ level, and population subsequently accumulates in Y1 due to spontaneous decay; after a delay $\tau_{\rm d}$, a weak frequency scan records the 1450\,nm absorption line. We use broadband scans that extend up to $\sim$3\,GHz, and 200\,MHz \emph{high-resolution sub-scans} for linewidth extraction. Read windows are of $\sim$1.2\,ms, and we use burn pulses down to 0.06\,ms.

\subsection{Theoretical Model (rate equations)}
\label{sec:model}
The three relevant manifolds are denoted by $\ket{g}$ (ground, $^3$H$_6$), $\ket{e}$ (optically pumped, $^3$H$_4$), and $\ket{b}$ (bottleneck, $^3$F$_4$), with populations $n_g$, $n_e$, $n_b$ and $n_g{+}n_e{+}n_b{=}1$. The pump excites $\ket{g}\!\to\!\ket{e}$ at an effective rate $R_e$ (ensemble-averaged over the addressed spectral fraction and optical depth). The $\ket{e}$ level decays with lifetime $T_1$ and branching ratio $\beta$ into $\ket{b}$ (and $1{-}\beta$ into $\ket{g}$). Note that the decay into $\ket{b}$ may also pass through the (short-lived) $^3H_5$ level, which is not  considered explicitly in the following. The $\ket{b}$ level relaxes to $\ket{g}$ with lifetime $T_b$. The rate equations are
\begin{align}
\dot n_g &= -R_e\,n_g + \frac{1-\beta}{T_1}\,n_e + \frac{1}{T_b}\,n_b, \nonumber\\
\dot n_e &= +R_e\,n_g - \frac{1}{T_1}\,n_e, \nonumber\\
\dot n_b &= \frac{\beta}{T_1}\,n_e - \frac{1}{T_b}\,n_b.
\label{eq:rates}
\end{align}
In steady state ($\dot n_i{=}0$) with $n_g{+}n_e{+}n_b{=}1$,
\begin{align}
n_g^{(\mathrm{ss})} &= \frac{1}{1 + R_e T_1 + \beta R_e T_b}, \qquad
n_e^{(\mathrm{ss})} = \frac{R_e T_1}{1 + R_e T_1 + \beta R_e T_b}, \qquad
n_b^{(\mathrm{ss})} = \frac{\beta R_e T_b}{1 + R_e T_1 + \beta R_e T_b}.
\label{eq:ss}
\end{align}
Only ions within the pump’s effective bandwidth $\Delta\nu$ are addressed. Denoting the inhomogeneous FWHM by $\Gamma_{\mathrm{inh}}$, an addressed fraction $f_{\rm spec}\!\approx\!\min(1,\Delta\nu/\Gamma_{\mathrm{inh}})$ scales the ensemble-averaged steady state. Optical-depth averaging over thickness $L$ with on-resonance absorption $\mathrm{OD}$ modifies the effective excitation rate as $\overline{R_e}=R_{e,0}(1-e^{-\mathrm{OD}})/\mathrm{OD}$. Power broadening increases $\Delta\nu$ via $s{=}I/I_{\rm sat}$.

For a weak probe, the \emph{integrated spectral hole area}\footnote{In spectral holeburning (SHB), the integrated area under a spectral hole is a fundamental quantity proportional to the optical transition dipole moment and directly related to the line intensity, providing a key parameter for understanding molecular properties and dynamics.} is proportional to the path-averaged bottleneck population:
\begin{align}
A(\tau_{\rm pump},\tau_{\rm d}) \propto \overline{n_b(\tau_{\rm pump},\tau_{\rm d})}.
\end{align}
Operationally,
\begin{align}
A(\tau_{\rm pump}) &\approx A_{\rm ss}\!\left[1 - a_1 e^{-\tau_{\rm pump}/\tau_1} - a_2 e^{-\tau_{\rm pump}/\tau_2}\right], \nonumber\\
A(\tau_{\rm d}) &\approx A(\tau_{\rm pump})\,e^{-\tau_{\rm d}/T_b}\qquad(\tau_{\rm d}\gtrsim 3T_1),
\end{align}
with $A_{\rm ss}\propto f_{\rm spec}\,n_b^{(\mathrm{ss})}$ and $\{\tau_1,\tau_2\}$ set by the $T_1/T_b$ hierarchy under the applied pump power.

\subsection{Simulations and Experiments}
\subsubsection*{Simulations and agreement with the rate-equation model}
The pump–wait–read protocol is simulated by numerically integrating Eqs.~\eqref{eq:rates} with steady-state relations in Eq.~\eqref{eq:ss}. The effective excitation rate $R_e$ is scaled to account for the finite pump bandwidth relative to the inhomogeneous line and for the averaging of optical depth along the sample. The simulated observables are the path-averaged bottleneck population $n_b(t)$ during pumping and its decay during the delay, converted to an \emph{integrated spectral hole area} $A$ for direct comparison with the measurements.

The simulations reproduce (i) a monotonic rise of $A(\tau_{\rm pump})$ towards $A_{\rm ss}\!\propto\!n_b^{(\mathrm{ss})}$ with a rapid approach set by $\ket{e}$ ($^3H_4$) and a slower approach limited by $\ket{b}$ ($^3F_4$); (ii) a single-exponential decay of $A(\tau_{\rm d})$ for $\tau_{\rm d}\!\gtrsim\!3T_1$, governed by $T_b$; and (iii) larger $A_{\rm ss}$ and broader spectral holes when the effective pump bandwidth was increased (by power broadening or deliberate scan width). The two eigenpolarisations at 795\,nm were taken into account by running the simulation at the two centers and comparing the resulting $A(\tau_{\rm pump})$ and FWHM trends.

The simulation curves were overlaid on the same axes used for the measurements (area versus pump duration, area versus delay, and linewidth versus power), resulting in quantitative agreement in functional forms and characteristic time scales. This agreement provides the baseline for extracting $T_1$, $T_b$, and preparation-dependent linewidths.

\subsubsection*{Experiments}
\label{sec:delft_experiments}
The protocol described above was implemented using the same Tm$^{3+}$:YAlO$_3$ crystal as in the main text, but at a slightly higher temperature of $\sim$4\,K. 

Two eigenpolarisations near 795\,nm (centers in $\sim$ 795.1\,nm and $\sim$ 795.3\,nm), corresponding to orthogonal crystal-axis orientations, were addressed to assess differences in the depth of the spectral hole, the line width and \emph{integrated spectral hole area} as functions of pump duration and power.

From pump–wait–read sequences with a subsequent delay, an excited state lifetime of $T_1\!\approx\!1$ ms for $\ket{e}$ and a bottleneck decay time of $T_b\!\approx\!6.3$–$6.4$\,ms for $\ket{b}$ were obtained, which are consistent with literature data and our alternative lifetime measurements described in the main text are detailed below. Representative scans of the 1450\,nm absorption line and fits (Fig.,S2) illustrate these values and the quality of the exponential fits. The spectral hole linewidth depends on preparation conditions: with a 0.06\,ms burn a narrow feature with FWHM $\sim$5.5\,MHz was observed; without magnetic field typical features near $\sim$10.5\,MHz were obtained; and with magnetic field applied the feature broadened to $\sim$19\,MHz. These measurements were consistent with the simulated functional forms and time scales employed for parameter extraction.\\

%\WT{We should add the branching ration $\beta$ and compare with literature ($\beta_{Lit.}\approx$ 16.5\%), where we remind the reader that our definition of $\beta$ corresponds to the sum of the branching ratios from X1 directly into Y1, and from X1 to the $^3H_5$ level, which subsequently decays rapidly into Y1. !!!!!!!CHECK!!!!!}

The rate-equation model also results in a branching ration of $\beta\!\approx\!17.2\%$, which is consistent with the literature value of 16.5$\%$ \cite{guillemot2019continuous}. Please note that this value describes the sum of the ratios for the decay from X1 directly into the $^3F_4$ manifold, and from X1 through the $^3H_5$ level into the $^3F_4$ manifold. In both cases, the population rapidly relaxes into the Y1 level.

%A branching ratio of $\beta\!\approx\!17.2\%$ for the excited state transition was also obtained using the rate equations model described above, which are also consistent with the literature value (16.5$\%$) \cite{guillemot2019continuous}. Here we define the branching ratio to be the sum of branching ratios from X1 to Y1 and from X1 to the $^3H_5$ level, which subsequently decays rapidly into Y1.}

Using this optical pumping scheme, the maximum optical depth we measured for the Y1 - X1 transition is $\sim$ 0.22. However, by directly pumping from Z1 into the \textsuperscript{3}F\textsubscript{4} manifold, we found a maximum optical depth of $\sim$ 1.12. Therefore, we chose direct optical pumping for the measurements described in the main text.

\section{Populating Y1 by direct pumping of population from Z1 to Y1}
In the main text we describe how to directly populate Y1 using a 1680 nm laser. We now turn our attention to the question of how to optimize the optical depth of the Y1 to X1 transition at around 1450 nm using such direct pumping. Towards this end, it is important to realize that the wavelengths of a specific ion for the Z1 to Y1 transition and for the Y1 to X1 transition are not correlated. Instead, pumping a spectrally narrow sub-ensemble of ions from Z1 into Y1 leads to a broad Y1 to X1 absorption line, whose width is given by the inhomogeneous linewidth of the latter transition and whose optical depth depends on the ratio between the bandwidth of the pump laser and the inhomogeneous Z1 to Y1 line. To optimize the optical depth of the 1450 nm Y1 to X1 transition, the spectral width over which the pumping takes place is therefore not relevant -- all that matters is the total number of ions that have been excited into the Y1 level. 
Due to the large mode-field diameter of the pump light inside the crystal, around 0.3 mm, the pump rate in our current setup is very small, and we are not able to empty the Z1 level, not even within the small bandwidth of the laser. We therefore opted to pump without additional frequency modulation of the laser, which would address ions with different homogeneous linewidth and therefore lead to a larger optical depth in the case in which we saturated the transition.

\section{Lifetime measurements}

\begin{figure}[h]
\includegraphics[width=0.95\linewidth]{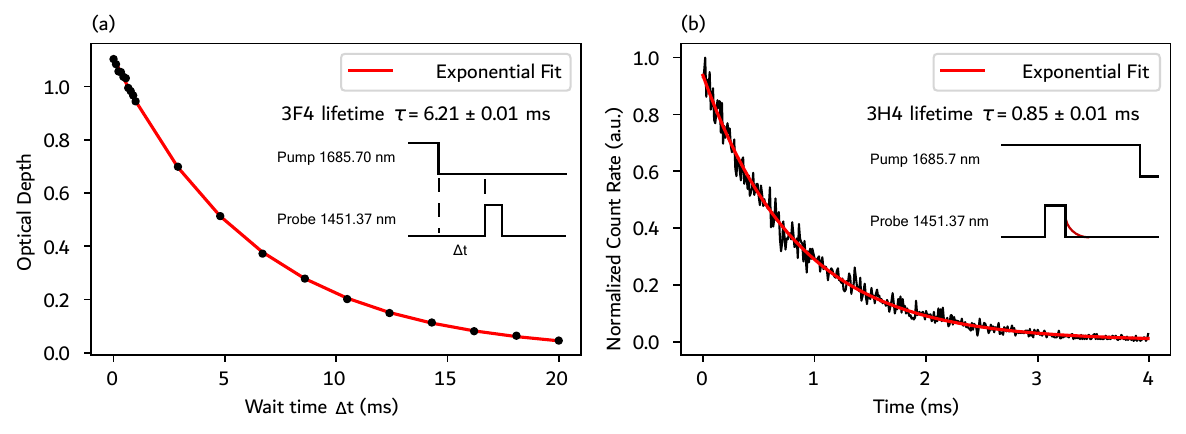}
\caption{\label{fig:lifetimes} (a) Time-resolved optical depth of the Y1-X1 transition at 1451.37 nm, measured at 3.16 K. (b) Photoluminescence decay at around 795 nm wavelength caused by X1 to \textsuperscript{3}H\textsubscript{6} transitions, measured at 1.04 K. }
\end{figure}

The decay of the optical depth of the Y1 - X1 transition reflects the decay of the ion population in Y1 within the \textsuperscript{3}F\textsubscript{4} manifold. In one measurement sequence, at first, the pump laser at 1685.70 nm is sent during $\sim$ 50 ms to populate Y1. After waiting for $\Delta t$, a probe pulse at 1451.37 nm is sent for 2 $\mathrm{\mu s}$, and the transmitted power is recorded. The waiting time $\Delta t$ is varied from 10 $\mathrm{\mu s}$ to 20 $\mathrm{ms}$. As shown in FIG. \ref{fig:lifetimes} (a), the decay of the optical depth, calculated using the recorded power levels, is well fitted by an exponential function, yielding a lifetime of Y1 of 6.21 ms. 

The lifetime of the X1 level within the \textsuperscript{3}H\textsubscript{4} manifold is  measured by monitoring photoluminescence. Towards this end, the pump laser at 1685.70 nm is kept constantly on, and a probe pulse at 1451.37 nm is sent for 200 $\mathrm{\mu s}$ to excite X1. The fluorescence signal at around 795 nm wavelength caused by spontaneous decay from X1 into the $^3$H$_6$ manifold can be detected right after the end of the probe pulse. The low-pass filter mentioned in the main text and in FIG. \ref{fig:experimental setup} ensures that photoluminescence due to any other decay is blocked. The exponential fit to the data shown in FIG. \ref{fig:lifetimes} (b) results in a lifetime of X1 of 0.85 ms. 

Note that the lifetimes of both the X1 and the Y1 levels align well with the results described in section \ref{sec:delft_experiments},  which were obtained using an entirely different procedure.

\section{Quadratic Zeeman shift}

As described in Eq. 1 of the main text, the magnetic field-dependent spin Hamiltonian of Tm energy levels contains a term that describes the level shifts caused by the quadratic Zeeman interaction.

\begin{figure}[h]
    \centering
    \includegraphics[width=0.5\linewidth]{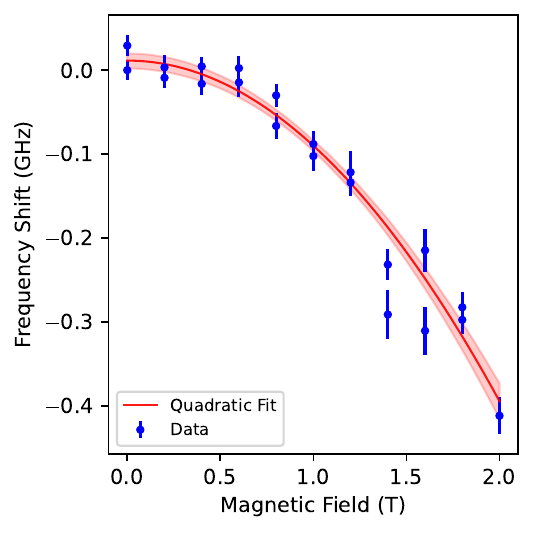}
    \caption{Frequency shift of the inhomogeneously broadened Y1 - X1 transition as a function of magnetic field}
    \label{fig:quad_zeeman_shift}
\end{figure}

Figure \ref{fig:quad_zeeman_shift} shows the frequency shift of the inhomogeneous broadened Y1 - X1 transition as a function of magnetic field. For each data point, we measured the frequency at the center of the line. The data are well-fitted by a quadratic function $\Delta f = a \cdot B^2 + c$, yielding $a = -101 \pm 3 \mathrm{MHz/T^2}$ and $c = 11 \pm 4 \mathrm{MHz}.$ The shaded region indicates the 95\% confidence interval of the fit. This measurement confirms the quadratic Zeeman interaction described in the spin Hamiltonian.

\section{Spectral hole burning}

\begin{figure}[h]
    \centering
    \includegraphics[width=0.6\linewidth]{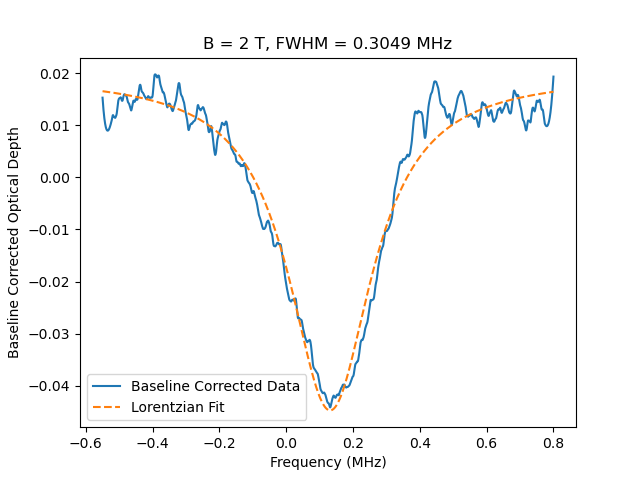}
    \caption{The narrowest measured spectral hole of the \textsuperscript{3}F\textsubscript{4} - \textsuperscript{3}H\textsubscript{4} ZPL.}
    \label{fig:the_narrowest_hole}
\end{figure}

The narrowest spectral hole we measured is shown in Figure \ref{fig:the_narrowest_hole}. The FWHM of this spectral hole is around 304.9 kHz, corresponding to an effective homogeneous linewidth of 66.95 kHz and hence T$_{2,eff}$=4.75 $\mu$s. The hole was measured when a 2 T external magnetic field was applied and at a temperature of 2.21 K. Optical pumping was done using the SLED, resulting in an optical depth of $\sim0.14$.

\subsection{Magnetic level splitting (Hyperfine splitting)}

%\subsection*{Hyperfine splitting}

\begin{figure}[h]
    \centering
    \includegraphics[width=0.9\linewidth]{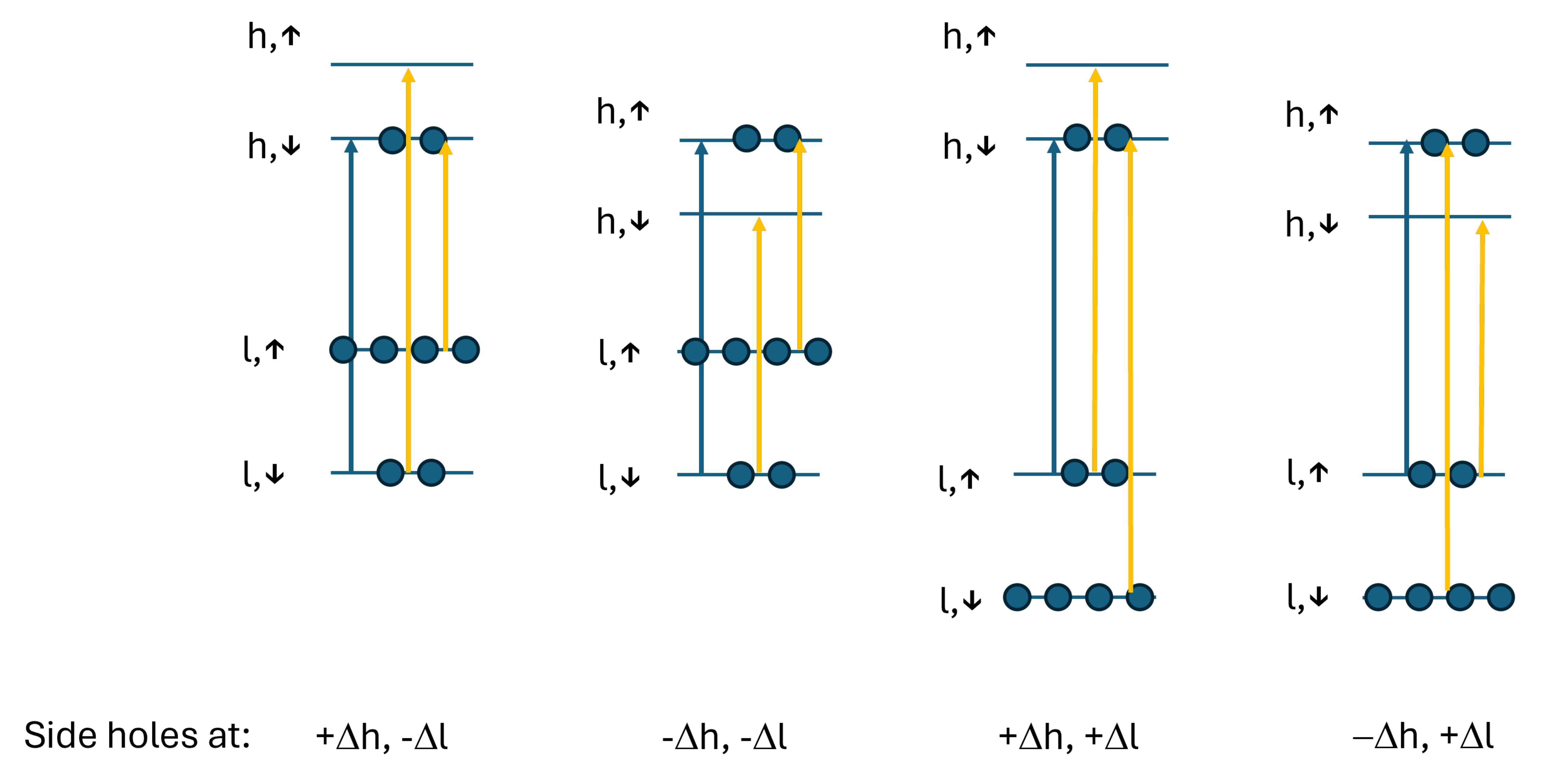}
    \caption{Hyperfine splitting in Tm:YAlO\textsubscript{3} and four scenarios (transitions) leading to population transfer into X1. Blue arrows denote the transitions driven by the burn pulse (of fixed frequency), and yellow arrows represent transitions (and frequencies) for which reduced optical depth (spectral holes) can be observed. The detunings at which sideholes (becoming clusters when taking superhypersplitting into account, as explained below) are observed are indicated below each scenario.}
    \label{fig:Hyperfine}
\end{figure}

As mentioned in the main text, when applying an external magnetic field, we observe two clusters (pairs) of side holes in the SHB measurements on either side of the central hole (these clusters can be distinguished from the central hole for B$>$0.4T). To explain this observation, let us denote the two magnetic hyperfine sublevels into which the \emph{lower} excited doublet Y1 splits as $(l,\downarrow)$ and $(l,\uparrow)$, and the two hyperfine sublevels that are part of the \emph{higher} excited doublet X1 as $(h,\downarrow)$ and $(h,\uparrow)$ (see FIG. \ref{fig:Hyperfine}). We furthermore define $\Delta l$ as the splitting of the Y1 doublet, and $\Delta h$ correspondingly as the splitting of the X1 doublet. Please note that in section \ref{sec:model} we use $b$ and $e$ instead of $l$ and $h$.

FIG. \ref{fig:Hyperfine} also depicts how, taking into account the inhomogeneous broadening of the Y1 - X1 transition, four different transitions can be driven by a burn pulse of a given frequency: ($l,\downarrow$) to ($h,\downarrow$), ($l,\downarrow$) to ($h,\uparrow$), ($l,\uparrow$) to ($h,\downarrow$), ($l,\uparrow$) to ($h,\uparrow$). 

Let us have a look at the first case, in which the burn pulse is resonant with the ($l,\downarrow$) to ($h,\downarrow$) transition. The central hole appears at the frequency of the burn pulse. But since the ion population in the levels ($l,\downarrow$) and ($h,\downarrow$) has been modified by the burn pulse, increased transparency (a side hole, or cluster) can be observed at the wavelength corresponding to the transition between ($l,\downarrow$) and ($h,\uparrow$), which is detuned by $+\Delta h$ from the central hole. In addition, we find another side hole (cluster) detuned by -$\Delta l$, corresponding to the transition between ($l,\uparrow$) and ($h,\downarrow$). We point out that this feature persists as long as the ($h,\downarrow$) level remains populated. This condition is satisfied in our experiments since the wait time between the burn and read pulse is only 10 $\mu$sec -- much shorter than the radiative lifetime of 0.85 ms of the ($h,\downarrow$) level. The latter also explains the lack of spectral intervals with increased absorption. Indeed, such anti holes can only be observed if the waiting time exceeded the radiative lifetime and spontaneous decay has led to an increased ion population in the $l,\uparrow$ level. In this case, anti holes appear at wavelengths corresponding to transitions between ($l,\uparrow$) and ($h,\downarrow$), and between ($l,\uparrow$) and ($h,\uparrow$), i.e. detunings of -$\Delta l$ and ($\Delta h$ - $\Delta l$).

Similar reasoning for the remaining three scenarios in FIG. \ref{fig:Hyperfine} allows us to conclude that there should be two clusters of side holes on either side of the central hole, with detunings of $\pm \Delta l$ and $\pm \Delta h$. This is consistent with our measured data.

\begin{figure}[htbp]
    \centering
    % First row
    \begin{subfigure}[t]{0.45\linewidth}
        \centering
        \includegraphics[width=\linewidth]{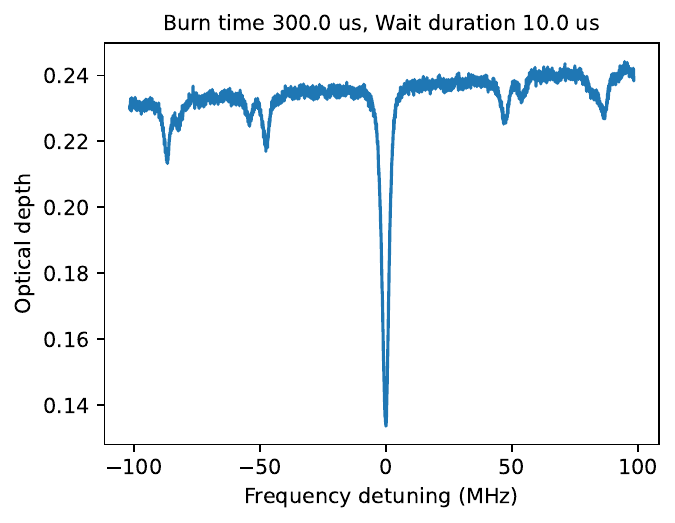}
        % \caption{Caption 1}
    \end{subfigure}
    \hfill
    \begin{subfigure}[t]{0.45\linewidth}
        \centering
        \includegraphics[width=\linewidth]{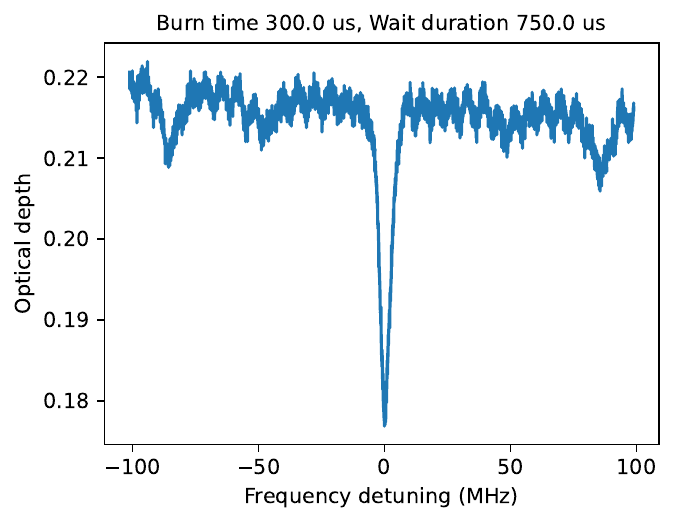}
        % \caption{Caption 2}
    \end{subfigure}

    \vspace{0.5cm} % space between rows

    % Second row
    \begin{subfigure}[t]{0.45\linewidth}
        \centering
        \includegraphics[width=\linewidth]{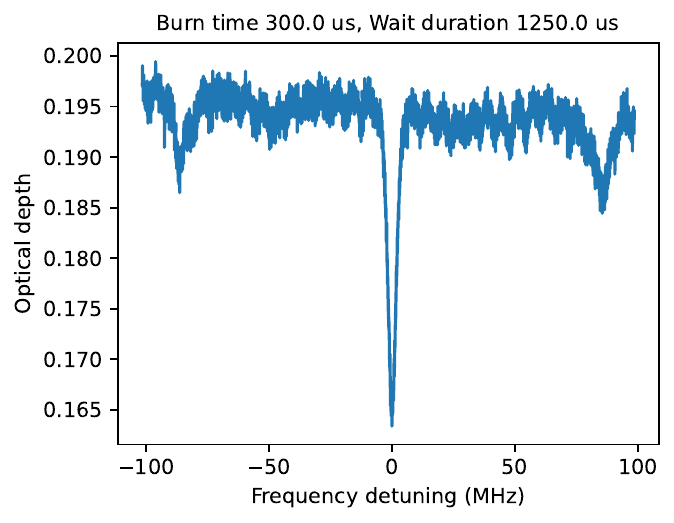}
        % \caption{Caption 3}
    \end{subfigure}
    \hfill
    \begin{subfigure}[t]{0.45\linewidth}
        \centering
        \includegraphics[width=\linewidth]{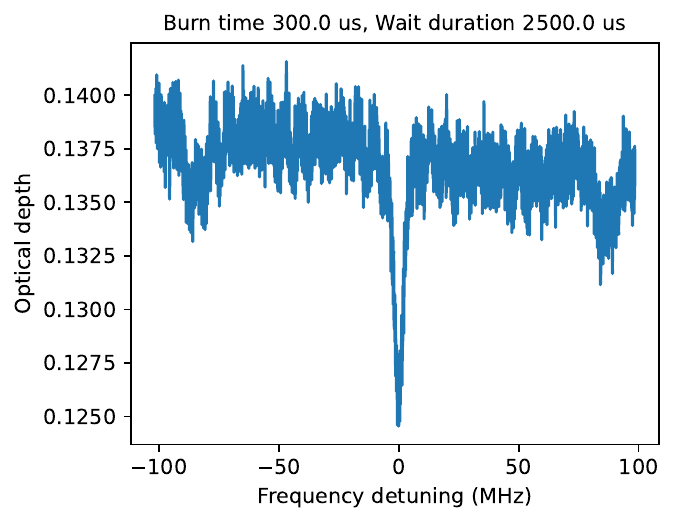}
        % \caption{Caption 4}
    \end{subfigure}
    \caption{Spectral hole burning with varying wait times : 10 $\mu$s, 750 $\mu$s, 1250 $\mu$s, and 2500 $\mu$s (clockwise starting at the top left figure).}
    \label{fig:time_resolved_SHB_varying_wait}
\end{figure}

\subsection{Time-resolved spectral hole burning}

The above model allows explaining the observation of two side holes (or clusters) on either side of the central hole. However, it does not allow identifying which of the observed splittings characterizes the splitting of the Y1 doublet $\Delta l$, and which one characterizes the splitting of the X1 doublet $\Delta h$. To remove this ambiguity, we perform another series of SHB measurements with an external magnetic field of 2T. We burn again for 300 $\mathrm{\mu s}$ but then vary the wait duration before we scan the laser over a spectral interval of $200 \mathrm{MHz}$ that is centered on the burn frequency. The results are depicted in FIG. \ref{fig:time_resolved_SHB_varying_wait}.

When the wait duration is set to 10 $\mathrm{\mu s}$, we clearly observe two clusters on either side of the central hole. The inner clusters (more precisely, the center of each cluster) feature a detuning of around $\pm 51 \mathrm{MHz}$, the outer clusters are at around $\pm 83 \mathrm{MHz}$. As the wait duration increases, we observe that the inner clusters decay with a lifetime comparable to the lifetime of the \textsuperscript{3}H\textsubscript{4} manifold (0.85 ms). This suggests that these clusters are due to a transition between the ``second" level within the Y1 doublet (not the level from which population is transferred by the burn laser) to the level within the X1 doublet into which population is pumped during burning. Using again the left-hand panel in FIG. \ref{fig:Hyperfine} as an example, the burn laser connects the  ($l,\downarrow$) level with the ($h,\downarrow$) level, and the inner clusters corresponds to a transition between the ($l,\uparrow$) level with the ($h,\downarrow$) level. Hence, the detuning of the inner {clusters with respect to the center hole reflects the splitting $\Delta l$ of the two hyperfine levels within the Y1 doublet.

The outer clusters decay at a rate that is comparable to that of the main hole. Similar reasoning leads to the conclusion that the frequency detuning of the outer clusters reflects the splitting $\Delta h$ of the two hyperfine levels within the X1 doublet.

\section{Free-induction decay}

Figure \ref{fig:pulse_sequence_FID} shows the pulse sequence used for optical FID measurements. A 50-ms-long pump pulse at 1685.7 nm is sent to populate Y1 within the \textsuperscript{3}F\textsubscript{4} manifold. Then a 20 $\mu$s long pulse at 1451.37 nm is sent to generate the FID signal. The latter is measured after the excitation pulse ends. 

\begin{figure}[h]
    \centering
    \includegraphics[width=0.5\linewidth]{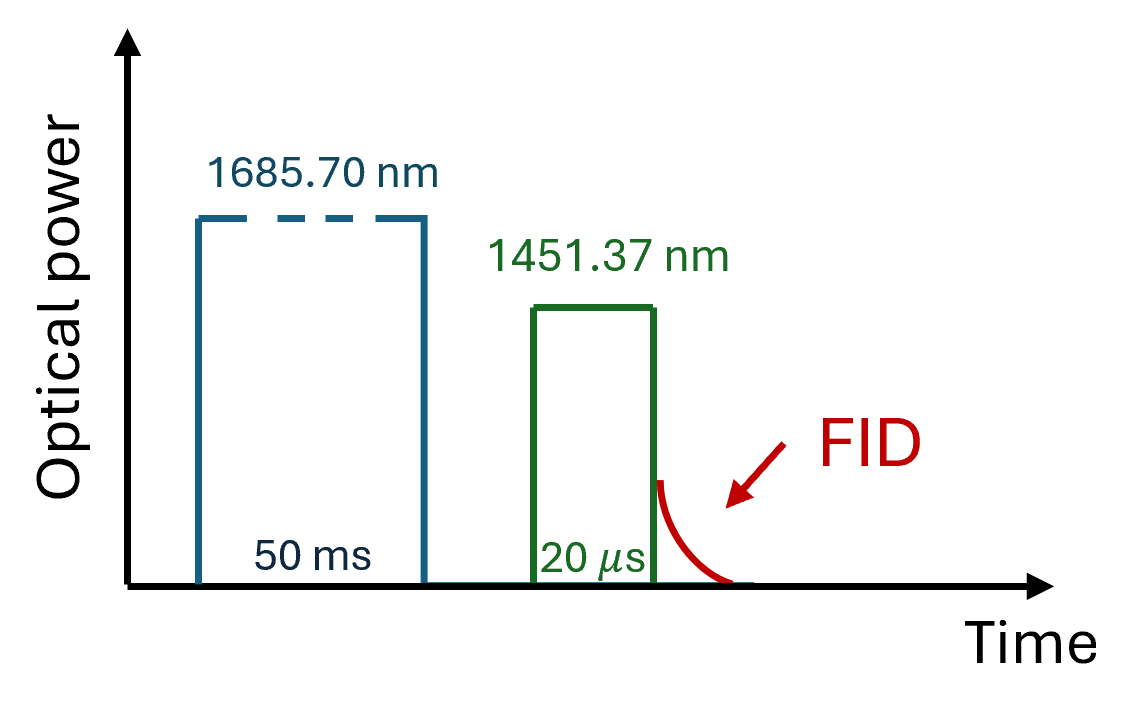}
    \caption{Pulse sequence for FID measurements.}
    \label{fig:pulse_sequence_FID}
\end{figure}

\subsection{FID measurements}
We use a photoreceiver (New Focus 1811) with a rise and fall time of 3 ns preceded by a gating AOM that cuts the pump and excitation pulse and prevents them from being detected. Figure \ref{fig:FID_measurements_varying_B_fields} shows some of FID measurements including exponential fits under different magnetic fields.

\begin{figure}[htbp]
    \centering
    % First row
    \begin{subfigure}[t]{0.45\linewidth}
        \centering
        \includegraphics[width=\linewidth]{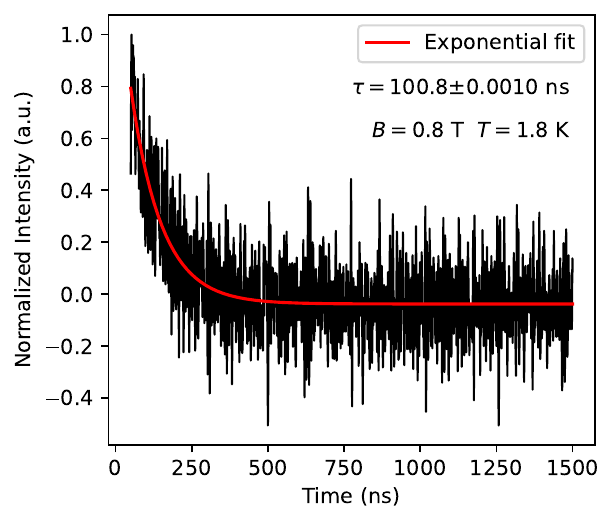}
        % \caption{Caption 1}
    \end{subfigure}
    \hfill
    \begin{subfigure}[t]{0.45\linewidth}
        \centering
        \includegraphics[width=\linewidth]{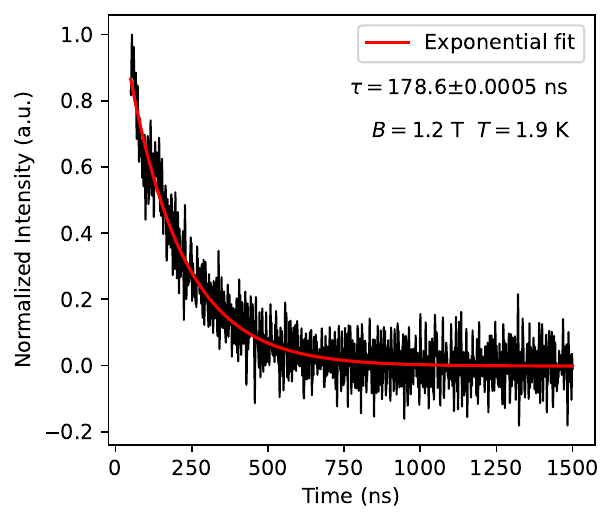}
        % \caption{Caption 2}
    \end{subfigure}

    \vspace{0.5cm} % space between rows

    % Second row
    \begin{subfigure}[t]{0.45\linewidth}
        \centering
        \includegraphics[width=\linewidth]{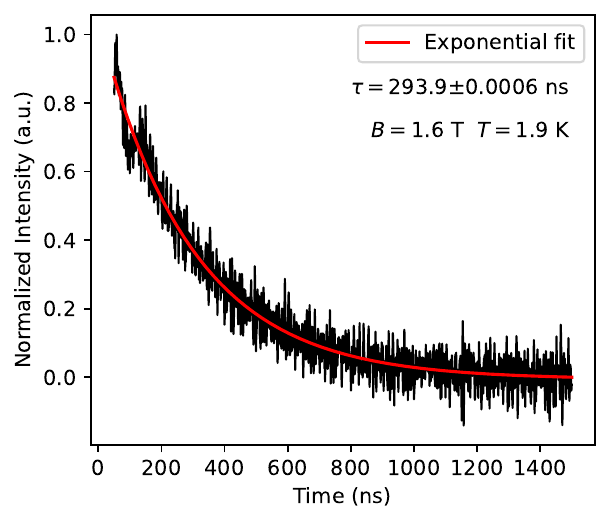}
        % \caption{Caption 3}
    \end{subfigure}
    \hfill
    \begin{subfigure}[t]{0.45\linewidth}
        \centering
        \includegraphics[width=\linewidth]{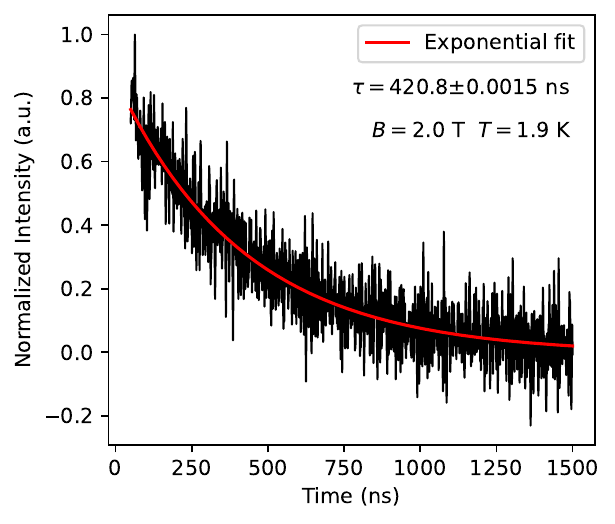}
        % \caption{Caption 4}
    \end{subfigure}
    \caption{Free induction decay with varying magnetic fields.}
    \label{fig:FID_measurements_varying_B_fields}
\end{figure}

\subsection{Derivation of optical coherence time as a function of FID decay time}

Following \cite{zangwill2013modern, milonni2010laser}, we consider a one-dimensional case of the propagation of the electromagnetic field in a material with light-matter interaction. From Maxwell equations in materials, with the slowly-varying amplitude approximation of a plane wave and the co-moving frame at the speed of light $c/n$, we can derive the following equation

\begin{equation}
    \frac{d E(t, z)}{d z}=i \frac{\mu_0 \omega c}{2 n} P(t, z),
\end{equation}
where $P(t, z)$ is the macroscopic polarization of the material. It acts as a source of the field emitted by the material.

If we consider the field propagates through a material with a short length of $\Delta z$, the field after $\Delta z$ can be approximated by
\begin{equation}
    E(t, z+\Delta z) = E(t, z) + i \frac{\mu_0 \omega c}{2 n} P(t, z),
\end{equation}
where $i \frac{\mu_0 \omega c}{2 n} P(t, z)$ is the field emitted by the material with length $\Delta z$.

We consider an ensemble of ions excited by an incoming pulse. The macroscopic polarization generated by this ensemble can be expressed as

\begin{equation}
    P(t, z)=N \int\langle\hat{\mu}(t, z, \delta)\rangle g(\delta) d \delta,
\end{equation}
where $N$ is the atomic density, $g(\delta)$ is the atomic density distribution of atoms that contributes to the macroscopic polarization, $\langle\hat{\mu}(t, z, \delta)\rangle$ is the expectation value of the electric dipole moment for a single atom at detuning $\delta$, and  $\delta$ is the detuning of the angular frequency. 

For a given atomic state $\hat{\rho}(t, z, \delta)$, the expectation value of the electric dipole moment is

\begin{equation}
    \langle\hat{\mu}(t, z, \delta)\rangle = \mathrm{tr}(\hat{\rho}(t, z, \delta)\hat{\mu}) \propto 2\mu_{ab} ( r_{1}(t, z, \delta) - ir_{2}(t, z, \delta)),
\end{equation}
where $r_{1}(t, z, \delta)$ and $r_{2}(t, z, \delta)$ are components of the Bloch vector of a two-level atomic state $\vec{r} = (r_{1}, r_{2}, r_{3})$.

The field emitted by this macroscopic polarization of an excited ensemble satisfies
\begin{equation}
    \frac{d E(t, z)}{d z} \propto \int\left[r_1(t, \delta)-i r_2(t, \delta)\right] g(\delta) d \delta.
    \label{eq:emitted field_1}
\end{equation}

The optical Bloch equations can be used to solve for $r_{1}(t, \delta)$ and $r_{2}(t, \delta)$ \cite{boyd2020nonlinear}. Let t = 0 be the end of the input excitation pulse, leading to initial conditions  $r_1(0)=0, r_{2}(0) = 1, \Omega(t) = 0$. We can derive
\begin{equation}
    r_1(t, \delta)=-e^{-\frac{t}{T_2}} \sin (\delta t),
    \label{eq:r1}
\end{equation}
\begin{equation}
    r_2(t, \delta)=e^{-\frac{t}{T_2}} \cos (\delta t).
    \label{eq:r2}
\end{equation}

Plugging eq.\ref{eq:r1} and \ref{eq:r2} into eq.\ref{eq:emitted field_1}, we find
\begin{equation}
    \frac{d E(t, z)}{d z}\propto e^{-\frac{t}{T_2}} \int e^{-i \delta t} g(\delta) d\delta.
    \label{eq:emitted_field_2}
\end{equation}

To calculate the integration in eq. \ref{eq:emitted_field_2}, we need to know the atomic density distribution $g(\delta)$ of the excited ensemble. As previously mentioned, the inhomogeneous broadening $\gamma_{\text{inh}}$ of our \textsuperscript{3}F\textsubscript{4} - \textsuperscript{3}H\textsubscript{4} ZPL is measured to be 1.29 GHz, which is orders of magnitude higher than the homogeneous linewidth $\gamma_{\text{hom}}$ and the linewidth of the laser $\gamma_{\text{laser}}$ that we used to generate the FID. Therefore, the atomic density distribution $g(\delta)$ satisfies
\begin{equation}
    g(\delta) \approx L_{\text{hom}}(\delta) \ast L_{\text{laser}}(\delta),
\end{equation}
where $L_{\text{hom}}(\delta)$ is the absorption spectrum of an homogeneous ensemble and $L_{\text{laser}}(\delta)$ is the laser spectrum. Both feature Lorentzian distributions. $g(\delta)$ is the convolution of both distributions, which is also described by a Lorentzian distribution. We denote the FWHM of $g(\delta)$ as $\Gamma_{g}$; it satisfies
\begin{equation}
    \Gamma_{g} = 2\pi\gamma_{g} \approx 2\pi(\gamma_{\text{hom}} + \gamma_{\text{laser}}),
    \label{eq:g_delta}
\end{equation}
where $\gamma_{\text{hom}}$ and $\gamma_{\text{laser}}$ denote linewidths in units of Hz.

Plugging eq.\ref{eq:g_delta} into eq.\ref{eq:emitted_field_2}, we notice that the integral describes the inverse Fourier transform of $g(\delta)$,

\begin{equation}
    \begin{aligned}
         \frac{d E(t, z)}{d z} & \propto e^{-\frac{t}{T_2}} \int e^{-i \delta t} g(\delta) d \delta \propto e^{-\frac{t}{T_2}} \frac{1}{2 \pi} \int e^{i \delta t} g(\delta) d \delta \\
        & \propto e^{-\frac{t}{T_2}} e^{-\frac{\Gamma_g}{2} t} \propto e^{-\pi \gamma_{\text {hom}} t} e^{-\pi\left(\gamma_{\text {hom}}+\gamma_{\text {laser}}\right)t} \\
        & \propto e^{-\pi\left(2 \gamma_{\text {hom}}+\gamma_{\text {laser}}\right)t}.
    \end{aligned}
\label{eq:emitted_field_3}
\end{equation}

Eq.\ref{eq:emitted_field_3} shows that the amplitude of the emitted field is proportional to an exponential function given by $e^{-\pi(2 \gamma_{\text{hom}}+\gamma_{\text{laser}})t}$. We define the free induction decay time $\tau_{\text{FID}}$ as the $1/e$ decay time of the intensity of the FID signal, which is half of the decay constant of the amplitude (as $I \propto |E^{2}|$). Hence,

\begin{equation}
    \tau_{\text{FID}} = \frac{1}{2\pi (2 \gamma_{\text {hom}}+\gamma_{\text {laser}})}.
    \label{eq:tau_FID}
\end{equation}

Using Eq.\ref{eq:tau_FID}, we can solve for the optical coherence time $T_{2}$, finding
\begin{equation}
    T_{2} = \frac{1}{\pi\gamma_{\text{hom}}} = \frac{4}{1/\tau_{\text{FID}} - 2\pi \gamma_{\text{laser}}}.
    \label{eq:T2_tau_FID}
\end{equation}
Eq.\ref{eq:T2_tau_FID} allows us to calculate the optical coherence time as a function of measured decay time of the free-induction decay.

\section{Spectral diffusion}

\subsection{Mechanism}

Spectral diffusion is the time-dependent perturbation of each ion’s transition frequency due to the dynamical nature of the ion’s environment. In the time domain, spectral diffusion accelerates decoherence whereas in the frequency domain, it causes the spectral hole to broaden over time. In our case, the local magnetic environment of each thulium ion is formed by the magnetic moments of the host (lattice) ions and the surrounding thulium ions. Whenever the spins of any of these ions change orientation, the local magnetic environment is perturbed, which results in spectral diffusion.

Following \cite{bottger2006optical, bottger2016optical}, we consider two mechanisms that lead to a change of spin orientation. One is phonon-driven spin flips, i.e. spin-lattice relaxation involving the absorption or emission of a  phonon that is resonant with the spin-flip transition. The other one is the spin flip-flop process, in which two nearby spins exchange energy through dipole-dipole interactions and simultaneously reverse their spin orientations.

As discussed in the main text, the broadening of the effective linewidth $\gamma_{eff}$ can be described by

 \begin{equation}
     \gamma_{\mathrm{eff}} = \gamma_{0} + \frac{\gamma_{\mathrm{SD}}}{2}[1 - e^{-Rt_{\mathrm{wait}}}],
     \label{eq:time_dependent_gamma_eff}
 \end{equation}
 where $\gamma_{0}$ is the intrinsic linewidth (no spectral diffusion), $\gamma_{\mathrm{SD}}$ is the maximum linewidth broadening introduced by spectral diffusion, and $R$ is the average bath spin-flip rate. Fig 3(a) in the main text demonstrates that we observe linewidth broadening introduced by spectral diffusion, with fitted average spin flip rate $R= 7.82\pm1.75 \text{kHz}$ and  $\gamma_{\mathrm{SD}} = 1041.7 \pm 119.1 \text{kHz}$.

\begin{figure}
    \centering
    \includegraphics[width=0.75\linewidth]{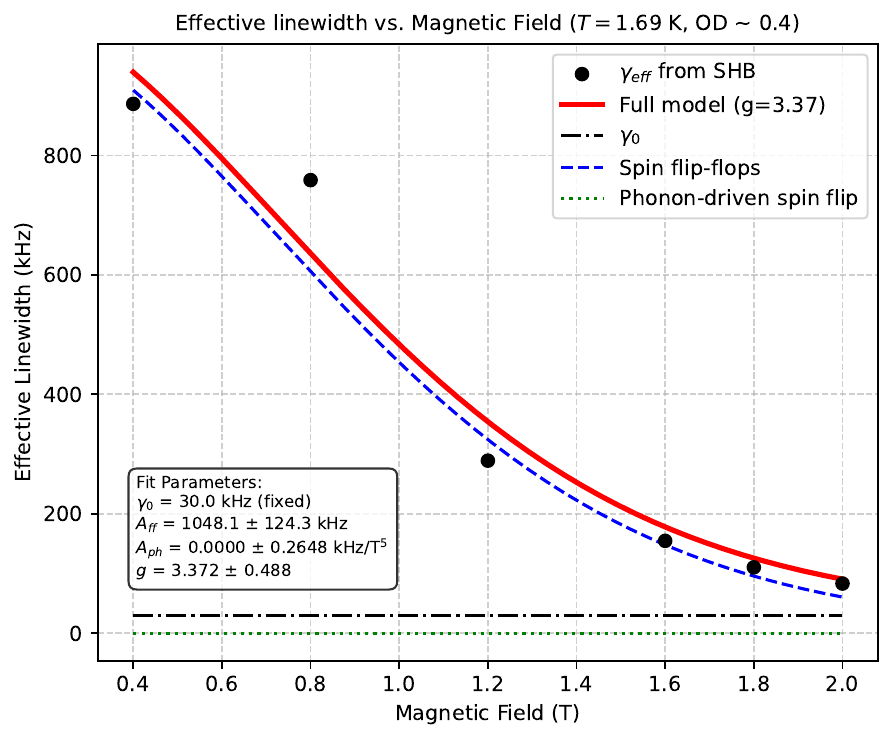}
    \caption{Effective linewidth measured by spectral hole burning with varying magnetic field, fitted by Eq. \ref{eq:gamma_eff_BT_full}.}
    \label{fig:SD_SHB}
\end{figure}

\begin{figure}
    \centering
    \includegraphics[width=0.75\linewidth]{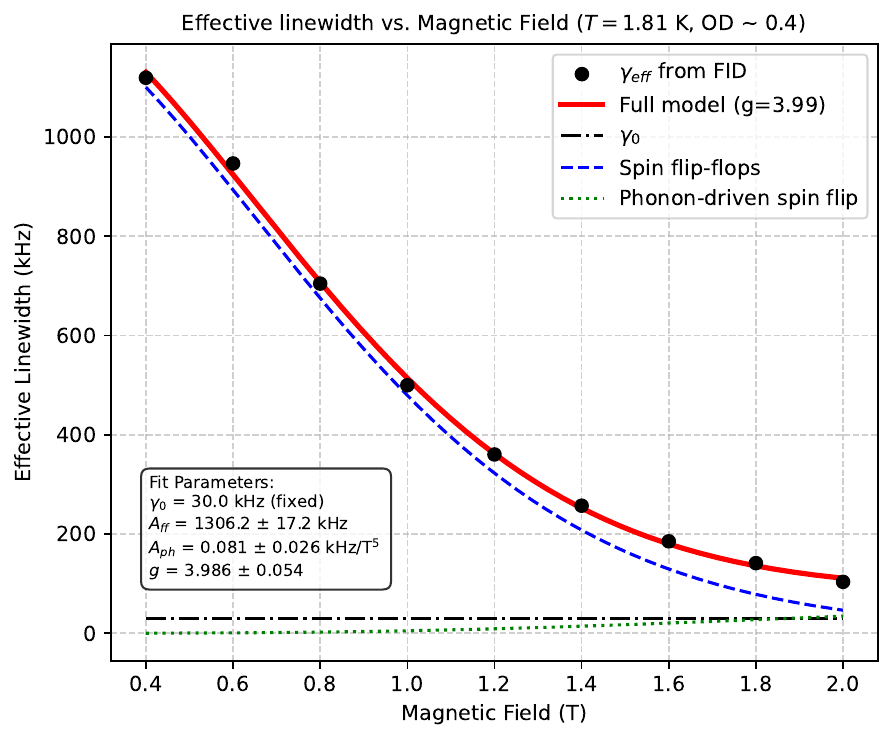}
    \caption{Effective linewidth measured by free induction decay with varying magnetic field, fitted by Eq. \ref{eq:gamma_eff_BT_full}.}
    \label{fig:SD_FID}
\end{figure}

We also investigated the magnetic field dependence of spectral diffusion. Both $R$ and $\gamma_{\mathrm{SD}}$ are dependent on the magnitude of the magnetic field. $\gamma_{\mathrm{SD}}$ is proportional to the number of spins that can undergo spin flips:

\begin{equation}
    \gamma_{\mathrm{SD}}(B) = \gamma_{\max} \,
    \operatorname{sech}^{2}\!\left( \frac{g_s \mu_B B}{2 k_B T} \right),
    \label{eq:gamma_SD}
\end{equation}
where $g_s$ is the effective gyromagnetic g factor of the paramagnetic spins along the direction of the applied field, $\mu_B$ is the Bohr magneton, $k_B$ is the Boltzmann constant, and $T$ is the temperature of the crystal.

Considering both spin flip-flops and phonon driven spin flips, the average spin flip rate can be written as

\begin{equation}
    R(B) = \alpha_{\mathrm{ff}} \,
    \frac{g_{\perp}^{2} \, n_s^{2}}{\Gamma_s} \,
    \operatorname{sech}^{2}\!\left( \frac{g_s \mu_B B}{2 k_B T} \right)
    + \alpha_{\mathrm{ph}} \, g_s^{3} \, B^{5} \,
    \coth\!\left( \frac{g_s \mu_B B}{2 k_B T} \right),
    \label{eq:R_full}
\end{equation}
where $g_{\perp}$ is the g tensor component of the paramagnetic spins perpendicular to the direction of the applied field B, $n_s$ is the spin density, and $\Gamma_s$ is the inhomogeneous broadening of the spin-flip transition \cite{seth2025spin}.

Plugging Eq.\ref{eq:gamma_SD} and Eq.\ref{eq:R_full} into Eq.\ref{eq:time_dependent_gamma_eff}, and assuming the wait time $t_{\mathrm{wait}} \ll 1/R$,  (which allows approximating $1-e^{Rt_{\mathrm{wait}}}$ with  $R\cdot t_{\mathrm{wait}}$), we can write the effective linewidth $\gamma_{\mathrm{eff}}(B)$ as

\begin{equation}
    \gamma_{\mathrm{eff}}(B) = \gamma_0
    + A_{\mathrm{ff}} \cdot
    \operatorname{sech}^{4}\!\left( \frac{g_s \mu_B B}{2 k_B T} \right)
    + A_{\mathrm{ph}} \cdot g_s^{3} \, B^{5}
    \operatorname{coth}\!\left( \frac{g_s \mu_B B}{2 k_B T} \right)
    \operatorname{sech}^{2}\!\left( \frac{g_s \mu_B B}{2 k_B T} \right).
    \label{eq:gamma_eff_BT_full}
\end{equation}

Figure \ref{fig:SD_SHB} and figure \ref{fig:SD_FID} show the effective linewidth measured by SHB and FID as a function of the magnetic field. Both data sets are fitted using Eq. \ref{eq:gamma_eff_BT_full}. We find in both cases that spin flip-flops are the main cause for spectral diffusion and are suppressed by increasing the magnetic field, and that the impact of phonon-driven spin flips is insignificant. Also, both sets of measurements give comparable values of the effective gyromagnetic g factor $g_{s}^{SHB} = 3.37\pm0.49$ and $g_{s}^{FID} = 3.99\pm0.05$.

\subsection{Possible sources of spectral diffusion}

The local magnetic environment of a doped Tm\textsuperscript{3+} ion in a YAlO\textsubscript{3} crystal is composed of other Tm\textsuperscript{3+} ions, of Al\textsuperscript{3+} ions with nuclear spin $I = 5/2$, and of Y\textsuperscript{3+} ions with nuclear spin $I = 1/2$. Tm\textsuperscript{3+} is a non-Kramers ion whose lowest crystal-field level is a singlet, and therefore carries no first-order electronic magnetic moment.

Previous research on YAlO\textsubscript{3} doped with non-Kramers ions attributed spectral diffusion to spin flip-flops between Al\textsuperscript{3+} nuclear spins \cite{bai1991spin, yano1992stimulated}. Al\textsuperscript{3+} has a nuclear g factor $g_n\approx1.457$, and its nuclear magneton is $\mu_{N}=5.051\times10^{-27} \mathrm{J/T}$. 
Limiting spectral diffusion to Al\textsuperscript{3+} spin flip-flops, we can calculate the line narrowing when changing the magnetic field from B=0T to B=2T. Assuming T=1.5K, we find

\begin{equation}
\begin{split}
    \gamma_{\mathrm{eff}}(B=2\mathrm{T}) &= \gamma_0
    + A_{\mathrm{ff}} \cdot
    \operatorname{sech}^{4}\!\left( \frac{g_n \mu_n B}{2 k_B T} \right) \\
    &\approx \gamma_0
    + A_{\mathrm{ff}} \cdot
    \operatorname{sech}^{4}\!\left( 0.00035 \right)
    \approx \gamma_0
    + A_{\mathrm{ff}} \cdot
    \operatorname{sech}^{4}\!\left( 0.0 \right)    
\end{split}
\label{eq:compare_eff_linewidth 0 and 2T}
\end{equation}
which is approximately equal to $\gamma_{\mathrm{eff}}$ at B=0T.
However, we measured an almost 10-fold reduction of the effective linewidth. This suggests that nuclear spin flip-flops between Al\textsuperscript{3+} ions do not cause the observed spectral diffusion in Tm\textsuperscript{3+}:YAlO\textsubscript{3}. And since the nuclear g-factors of Y\textsuperscript{3+} and Tm\textsuperscript{3+} are even smaller than that of Al\textsuperscript{3+}, they can be expected to contribute even less to spectral diffusion.

However, we note that the energy gap between the two lowest Stark levels Z1 and Z2 in the \textsuperscript{3}H\textsubscript{6} manifold is very small ($\sim 90 \mathrm{GHz}$). By applying an external magnetic field, these two levels are mixed and form a non-Kramers quasi-doublet that exhibits an electronic magnetic moment through the Van Vleck paramagnetism \cite{van1932theory}. As we increase the magnetic field, the energy gap between Z1 and Z2 increases due to the quadratic Zeeman interaction. Consequently, the electronic magnetic moments (quasi-doublets) bath is further polarized into the low-energy state. Therefore, the resonant antiparallel pairs, which are the source of spin flip-flops, become rarer \cite{bottger2006optical}, leading to less significant spectral diffusion.

\subsection{Prediction of optical coherence at lower temperatures}

\begin{figure}[h]
    \centering
    \includegraphics[width=0.75\linewidth]{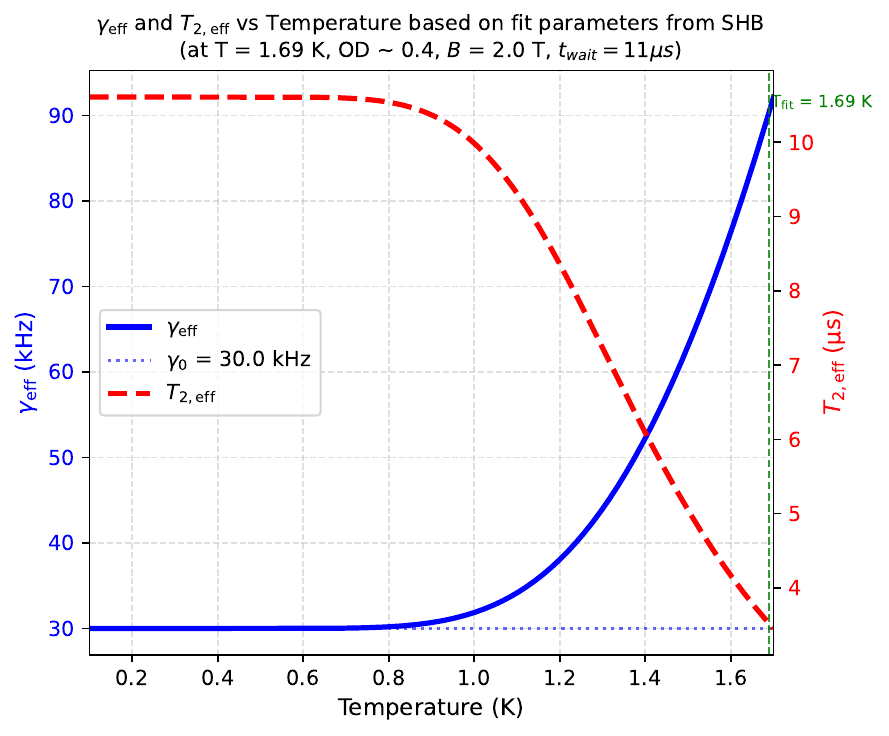}
    \caption{Prediction of effective linewidth and effective optical coherence time as a function of temperature using  Eq.\ref{eq:gamma_eff_BT_full} and the fitted parameters shown in Fig. \ref{fig:SD_SHB}. The optical depth in these measurements was 0.4.}
    \label{fig:SD_temp_dependent}
\end{figure}

The spectral diffusion model (Eq.\ref{eq:gamma_eff_BT_full}) along with the fitted parameters from Fig. \ref{fig:SD_SHB} allows us to predict the increase of $T_{2,\text{eff}}$ as we decrease the sample temperature. Fig. \ref{fig:SD_temp_dependent} shows that  $T_{2,\text{eff}}$ saturates at around $10.6 \mu s$ for T$<$0.8K, which is consistent with the value of the intrinsic linewidth $\gamma_0$ =30 kHz. Note that this prediction is based on measurements with an optical depth (OD) of 0.4 --  it is reasonable to expect a longer optical coherence when decreasing the OD.

\section{Towards quantum technology}
In this section we briefly discuss the potential for the 1450 nm transition of Tm:YAlO$_3$ for creating quantum network technology, more precisely quantum memory for light, and single photon sources.

\subsection{Optical quantum memory} 
A widely explored approach to quantum memory for light is based on ensembles of rare-earth ions doped into transparent crystals and the so-called atomic-frequency-comb memory \cite{afzelius2009multimode,tittel2025quantum}. After an initial spectral tailoring step that shapes the inhomogeneously broadened absorption line into a series of narrow and equally spaced absorption teeth, a single photon can be absorbed and later re-emitted with high fidelity. The storage time is given by the inverse tooth spacing and can be further increased by mapping the optically excited atomic coherence reversibly onto longer-lived spin coherence. To ensure absorption of the photon, recent developments enhance the light-atom interaction using an impedance-matched cavity \cite{Afzelius2010}. This has allowed increasing the storage efficiency, e.g. from  below 1\% for a crystal with single-pass optical depth after spectral tailoring of 0.4 (which would be similar for the Tm:YAlO$_3$ crystal used in our demonstration) to 27.5\% \cite{Davidson2020}. Note that this efficiency was limited by imperfect impedance matching and imperfect spectral tailoring, and that the state-of-the-art has meanwhile reached 80\% \cite{Meng2025}.

\begin{figure}[h]
    \centering
    \includegraphics[width=0.95\linewidth]{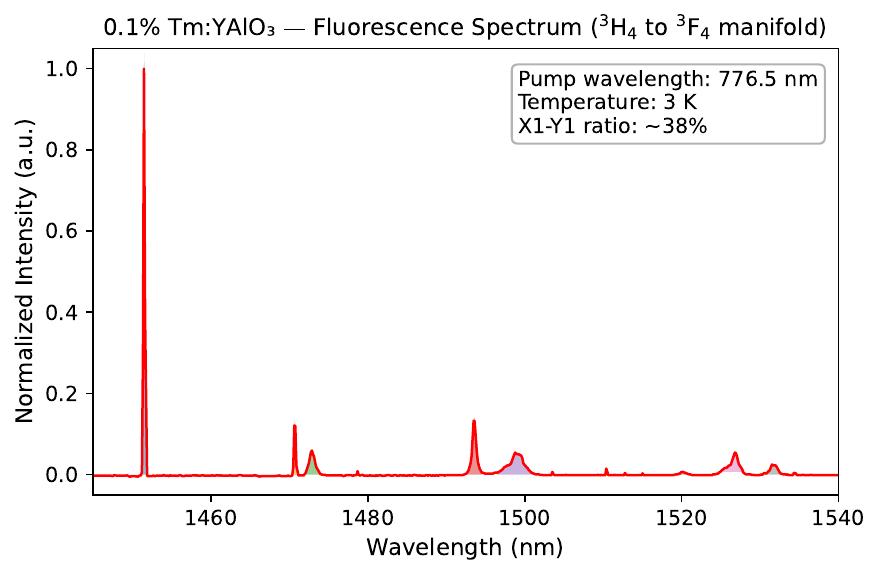}
    \caption{Fluorescence spectrum for decay from X1 to any crystal field level within the \textsuperscript{3}F\textsubscript{4} manifold after exciting the $^3H_4$ manifold using horizontally polarized light of 776 nm wavelength. We estimate the probability for the X1 to Y1 transition to be around 38\%.} 
    \label{fig:branching}
\end{figure}

\subsection{Single-photon sources}
As has been demonstrated by several groups in the world (see \cite{tittel2025quantum} for a recent review), it is possible to observe single photons from individual rare-earth ions using Purcell-enhanced light-matter interaction. This requires coupling the rare-earth ion to the optical mode of a cavity with high quality factor $Q$ and small mode volume $V_{mode}$. Assuming for simplicity that the cavity resonance and the ion resonance coincide and that the polarization of the latter is optimized, the Purcell factor $F_P$ (which determines the increase of the photon emission rate from $\gamma_0$ to $\gamma'$, with $\gamma'=F_P\gamma_0$) is given by
\[
F_P=\frac{3}{4\pi^2}\left (\frac{\lambda}{n}\right )^3\frac{Q}{V_{mode}}\abs{\frac{E(r)}{E_{max}}}^2\cdot\frac{\beta}{\chi_L}.
\]
Here, $\lambda$ is the resonance wavelength, $n$ the refractive index of the rare-earth crystal, and $E(r)$ and $E_{max}$ the electric field at the position of the ion and the maximum electric field inside the cavity, respectively. Furthermore, $\beta$ is the branching ratio for the desired transition, in our case from X1 to Y1, and $\chi_L=[(n^2+2)/3]^2$ the local field correction \cite{yu2023frequency}. Using $\lambda$=1450 nm, $n$=1.92, and Q=50 000,  V=0.09$\mu m^3$ and $E(r)/E_{max}=0.1$, as in \cite{yu2023frequency}, we find 
\[F_P\approx 2200\cdot\beta.
\]
We can estimate the branching ratio $\beta$ from the branching ratio of 11.3\% that describes the probability that an ion in state X1 decays radiatively into \emph{any} crystal field level within the \textsuperscript{3}F\textsubscript{4} manifold \cite{guillemot2019continuous}, multiplied with the probability that the decay terminates in Y1. Estimating the latter from Fig. \ref{fig:branching} to be around 38\%, we find 
\[
\beta\approx 0.04
\]
and hence 
\[
F_P\approx 90.
\]
This value can be further increased using a cavity with larger Q/V \cite{Seidler2013}, and/or by creating the cavity directly out of Tm:YAlO$_3$, in which case $E(r)/E_{max}$ can in principle reach 1. The creation of single-photon sources based on the 1450 nm excited-state transition in Tm:YAlO$_3$ therefore appears feasible. 

\bibliography{references}